\documentclass[pdftex,twocolumn,epjc3]{svjour3}          
\usepackage[english]{babel}
\RequirePackage{graphicx}
\usepackage{graphicx,amsmath,amssymb,xcolor,cite,multirow}

\journalname{Eur. Phys. J. C}

\newcommand{\Schw}{Schwarzschild}

\newcommand{\beq}{\begin{equation}}
\newcommand{\eeq}{\end{equation}}
\newcommand{\bea}{\begin{eqnarray}}
\newcommand{\eea}{\end{eqnarray}}
\newcommand{\non}{\nonumber}

\newcommand{\aaa}{{\diamond}}
\newcommand{\ce}{{\cal{E}}} \newcommand{\cl}{{\cal{L}}} \newcommand{\cb}{{\cal{B}}}

\def\p{p}  \def\cp{\pi}
\def\x{x}

\def\mJ{I} \def\mD{II}

\def\af{\zeta}

\def\d{\dif}

\def\xm{w} 

\def\BB{{\cal B}}

\def\tt{\theta}

\def\af{\zeta}

\providecommand{\dif}{\mathrm{d}} \def\d{\dif}

\def\mir{\mathrm{r}}
\def\mit{\mathrm{\theta}}
\def\mip{\mathrm{\phi}}

\graphicspath{{pic/}}

\par
\par

\begin{document}

\title{
Charged particle dynamics in parabolic magnetosphere around Schwarzschild black hole
}

\author{
Martin Kolo\v{s}\thanksref{e1,addr1}
\and
Misbah Shahzadi\thanksref{e2,addr2}
\and
Arman Tursunov\thanksref{e3,addr1}
}
\thankstext{e1}{e-mail: martin.kolos@physics.slu.cz}
\thankstext{e2}{e-mail: misbahshahzadi51@gmail.com}
\thankstext{e3}{e-mail: arman.tursunov@physics.slu.cz}

\institute{Research Centre for Theoretical Physics and Astrophysics, Institute of Physics, Silesian University in Opava, Bezru{\v c}ovo n{\'a}m.13, CZ-74601 Opava, Czech Republic.\label{addr1}
\and
Department of Mathematics, COMSATS University Islamabad, Lahore Campus, 54000 Lahore, Pakistan.\label{addr2}
}


\date{Received: date / Accepted: date}
\maketitle

\begin{abstract}
The study of charged particle dynamics in the combined gravitational and magnetic field can provide important theoretical insight into astrophysical processes around black holes. In this paper, we explore the charged particle dynamics in parabolic magnetic field configuration around Schwarzschild black hole, since the paraboloidal shapes of magnetic field lines around black holes are well motivated by the numerical simulations and supported by observations of relativistic jets. Analysing the stability of bounded orbits and using the effective potential approach, we show the possibility of existence of stable circular off-equatorial orbits around the symmetry axis. We also show the influence of radiation reaction force on the dynamics of charged particles, in particular on the chaoticity of the motion and Poincar\'{e} sections, oscillatory frequencies, and emitted electromagnetic spectrum. Applied to Keplerian accretion disks, we show that in parabolic magnetic field configuration, the thin accretion configurations can be either destroyed or transformed into a thick toroidal structure given the radiation reaction and electromagnetic-disk interactions included. Calculating the Fourier spectra for radiating charged particle trajectories, we find that the radiation reaction force does not affect the main frequency peaks, however, it lowers the higher harmonics making the spectrum more flat and diluted in high frequency range. 
%
\end{abstract}

\date{\today}



\section{Introduction} \label{intro}



Long-range forces provided by gravitational and electromagnetic (EM) interactions are of crucial importance for the proper understanding of high-energy processes around black holes (BHs). There are convincing observational evidences that magnetic fields (MFs) must be present in the vicinity of BHs \cite{Eatough-etal:2013:Natur:,Daly:2019:APJ:}. Orders of magnitude of MFs around BHs may vary from a few Gs up to $10^8$~Gs and more, depending on the source generating the field. For stellar-mass BHs observed in X-ray binaries, the characteristic strength of MFs is of the order of $10^8$~Gs, while for supermassive BHs (SMBHs), the characteristic strength is of the order of $10^1-10^4$~Gs \cite{Eatough-etal:2013:Natur:,Gold-etal:2017:ApJ:,EHT:2021:ApJ:}. Since the energy densities of MFs of such orders are not enough to make a sufficient contribution to the geometry of the background spacetime, in realistic astrophysical situations the spacetime metric around a BH can be fully described by the Kerr or \Schw{} solution of the Einstein field equations.

The weakness of the MF around BH, in the sense that the MF does not affect the spacetime geometry, is compensated by the large value of the charge-to-mass ratio for elementary particles, whose motion will be essentially affected by MFs already of the order of few Gs. For a charged test particle with charge $q$ and mass $m$ moving in the vicinity of a BH with mass $M$ immersed in MF of the strength $B$, one can introduce a dimensionless ``magnetic parameter'' $\cal{B}$ reflecting the ratio between the Lorentz force (LF) and gravitational force. For a relativistic electron orbiting a BH at the distance of the event horizon scale, one can estimate this ratio as follows
\beq
{\cal{B}}  \sim B \, \frac{ q }{ m } \, \frac{ G M }{ c^4 } \approx \, 10^{11} \,\, \frac{q}{e} \, \frac{m_{\rm e}}{m} \, \frac{B}{10^8 {\rm G}} \, \frac{M}{10 M_{\odot}}  ,
\eeq
that is very large due to large values of the specific charge $q/m$ for elementary particles. Therefore, the influence of the EM force on the motion of charged particles can be dominant in realistic situations.

\begin{figure}
\centering
\includegraphics[width=0.9\hsize]{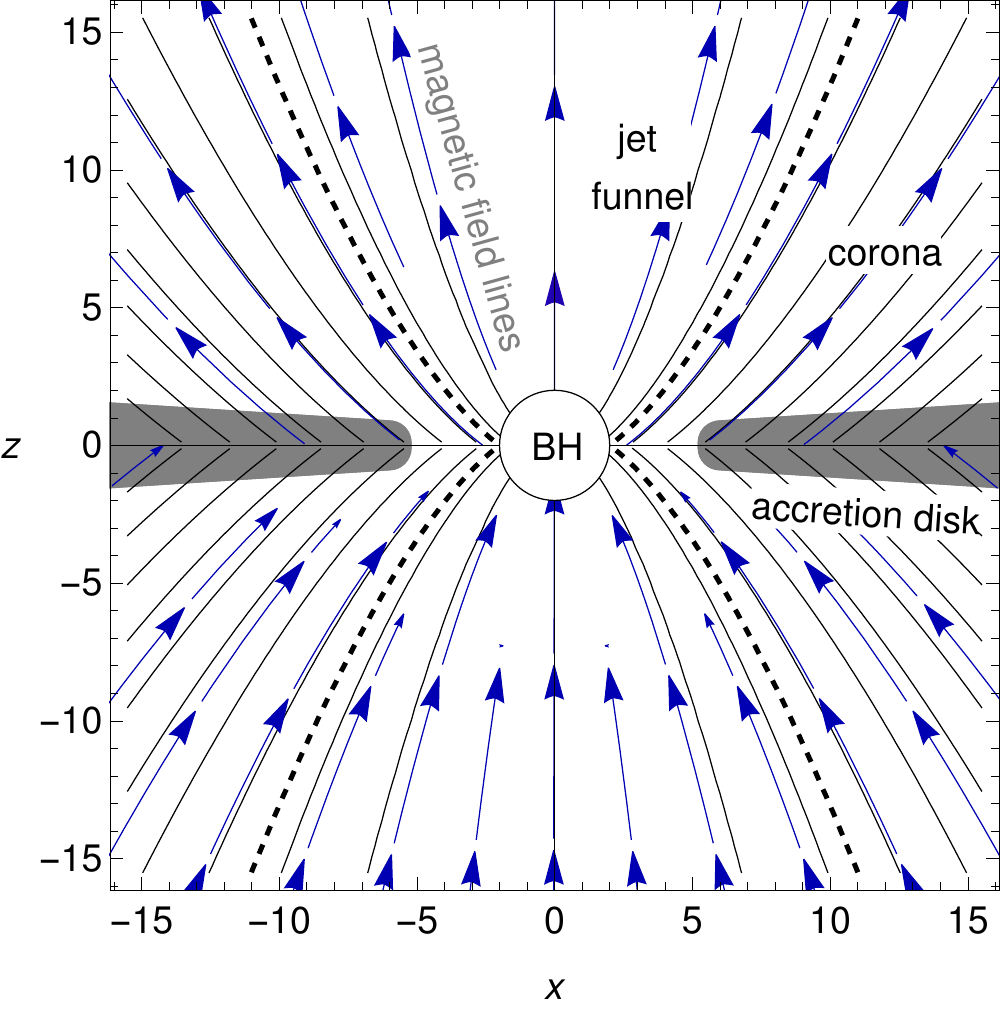}
\caption{
Parabolic MF configuration around Schwarzschild BH. Quasi neutral Keplerian accretion disk formed by particles on circular orbits close to the equatorial plane is restricted in its inner edge by the equatorial innermost stable circular orbit (inISCO). The dashed curve is the boundary between the accretion disk corona and jet funnel region close to the $z$-axis.
\label{fig_motiv}
}
\end{figure}


Within the theory of general relativity, the gravitational influence of a central compact object (BH) can be well described by the Kerr or \Schw{} spacetime \cite{Stu-etal:2020:Universe:}, but unfortunately, we don't have such an elegant and well accepted solution for EM field influence -- instead, there exist various models for BH magnetosphere depending on the complexity of included physical effects. 
As a zero approximation to the problem of BH magnetosphere one can use the vacuum solutions of Maxwell equations in the curved background as a starting point. In such approximation there is no source $\rho=0$, or EM currents $J^\alpha=0$ included, and the solution can be considered as an EM field of isolated BH in vacuum. 
However, electrovacuum solutions have limited astrophysical relevance, as they do not include the effects of material orbiting around BH in the form of a plasma. In the force-free approximation plasma is taken into account, but MF energy density is still dominating over matter energy density $B^2\gg\rho c^2$. In such models, the matter is used to generate EM currents inside the BH magnetosphere, but the matter is still locked in the MF. Plasma effect on the force-free magnetosphere of BH has been studied in the well-known work of Blandford \& Znajek \cite{Bla-Zna:1977:MNRAS:}, where also the EM mechanism of  rotational energy extraction from BH has been proposed. 
Several numerical techniques have been also employed, but the exact shape and intensity of the BH magnetosphere are not yet properly resolved, although a strong connection to the accretion processes is evident \cite{Meier:TheEngineParadigm,Galishnikova-etal:2022:XXX:}. One can also use the general relativistic magnetohydrodynamics (GRMHD) or general-relativistic particle-in-cell (GRPIC) numerical simulations to test EM field configurations and try to construct a relevant BH magnetosphere model.  A simple and still relevant BH magnetosphere model inspired by GRMHD and GRPIC simulations is the parabolic MF configuration \cite{Nak-etal:2018:APJ:,Por-etal:2019:APJS:,Kol-Jan:2020:RAG:,Cri-etal:2020:AA:}.


MF effects on the astrophysical phenomena occurring in the vicinity of BHs cannot be neglected and the aim of this article is to find close connections between a simple realistic model for the BH magnetosphere with a radiative processes in the combined gravitational and EM fields. Simple charged particle model has limited validity in description of matter around BH -- one should assume low particle density, i.e., that the mean time of two particle interaction is longer than orbital time around BH. One should also note that the typical Larmor radius of a particle in BH magnetosphere is much smaller than the Schwarzschild radius. Such low particle density assumption is correct only for the jet and for disks of low-luminosity active galactic nuclei (AGN), like Sgr~A* or M87, while plasma inside thin disk around luminous AGN or XRBs may be quite collisional due to thin accretion disk's high density.

Our work can be seen as preliminary basic study of realistic BH magnetosphere, which should have according to the recent observation of synchrotron radiation and numerical modelling parabolic character \cite{Nak-etal:2018:APJ:}. The article is structured as follows. We explore charged particle motion in parabolic BH magnetosphere model using effective potential approach in section \ref{SecVEFF}, then we examine the influence of radiation reaction (RR) damping force on particle chaotic dynamics in section \ref{SecCHAOS}. We tested neutral Keplerian disk ionisation in section \ref{secDisk}, estimate observable frequencies in section \ref{oscillations}, and provide some astrophysical application in section \ref{SecASTRO}.

Throughout the paper, we use the spacelike signature $(-,+,+,+)$. Greek indices are taken to run from $0$ to $3$, while Latin indices run only over the spatial coordinates ($r, \theta, \phi$). However, for expressions having astrophysical relevance, we use the physical constants explicitly.

\begin{figure*}
\includegraphics[width=0.47\hsize]{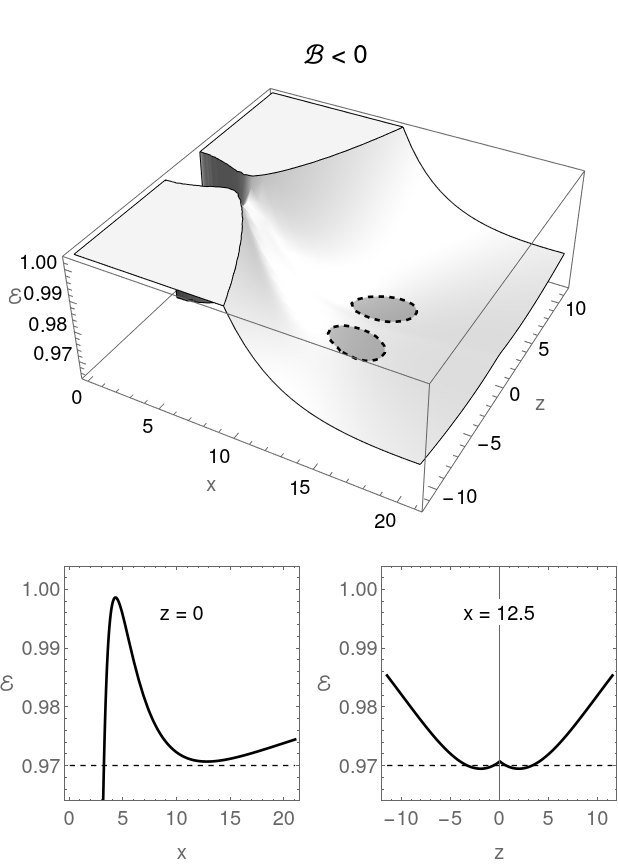}
\includegraphics[width=0.47\hsize]{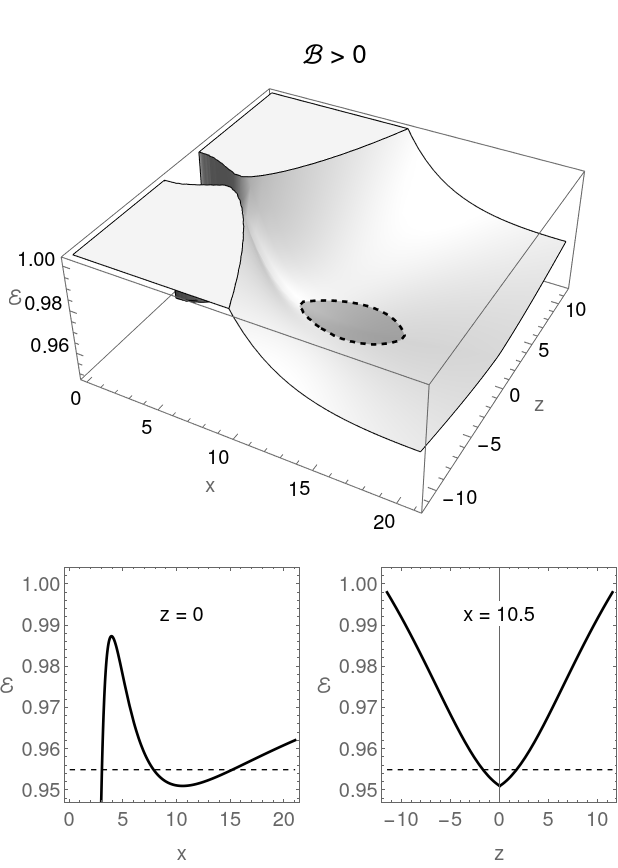}
\caption{Effective potential $V_{\rm eff}$ for charged particle dynamics as 2D function of Cartesian $x,z$ coordinates. We can see the effective potential minima at the off-equatorial planes for $\cb<0$ (top left) and the in-equatorial minima for $\cb>0$ (top right). The lower row of figures shows the sections of the effective potential at given energy and the corresponding minima. Value of the parameter $\xm$ has been set to $\xm=3/4$, but similar behavior is observed for different values of $\xm \in(0,1)$.
\label{figVeff}
}
\end{figure*}

\section{Charged particles in black hole magnetosphere} \label{SecVEFF}

The line element describing the spacetime of non-rotating Schwarzschild BH with mass $M$ is given as
\beq
    \d s^2 = -f(r) \d t^2 + f^{-1}(r) \d r^2 + r^2(\d \theta^2 + \sin^2\theta \d \phi^2), \label{SCHmetric}
\eeq
where the function $f(r)$ takes the form
\beq
	f(r) = 1 - \frac{2 M}{r}.
\eeq
Hereafter, we put $M=1$, i.e., we use dimensionless radial coordinate $r$ (and time coordinate $t$).

In our magnetosphere model, we  assume the axial symmetry and absence of electric field, hence the only non-zero component of EM four-vector potential $A_{\mu}$ will be $A_\phi$, so that $ A_\mu = (0,0,0,A_\phi)$. We will consider heuristic BH magnetosphere model given by the parabolic MF solution motivated by the GRMHD plasma simulations \cite{McK-Nar:2007A:MNRAS:,McK-Nar:2007B:MNRAS:,Tch-Nar-McK:2010:APJ:,Cri-etal:2020:AA:}
\beq
A_{\phi} (r,\theta) = \frac{B}{2} \, r^{\xm} \, (1- | \cos\theta | ), \label{mag_field}
\eeq
where $B\in(-\infty,\infty)$ is MF intensity and $\xm\in[0,1.25]$ is declination of MF lines. Such parabolic field configuration should represent magnetic field inside BH jet region (close to $z$-axis) and more complex configuration is expected in accretion disk region (close to equatorial plane). One should note that (\ref{mag_field}) is  
a simplified version of the Blandford-Znajek (BZ) paraboloidal model \cite{Bla-Zna:1977:MNRAS:} and it is not a solution of a vacuum Maxwell equation in \Schw{} spacetime. The absolute value in $A_{\phi}$ definition is making the solution to split in equatorial plane and preserving MF divergence, so MF lines go into the BH below the equatorial plane and outgoing above the equatorial plane, see Fig.~\ref{fig_motiv}. Such split magnetic configuration would require existence of conductive plasma current sheet in equatorial plane, thus a material orbiting around and falling into BH is supporting this model. In the case with $\xm=1$ one gets a limit of BZ paraboloidal model and for  $\xm=0$ the BZ split monopole solution \cite{Bla-Zna:1977:MNRAS:}. The value, which is mostly used for BH magnetosphere in jet funnel is $\xm=3/4$ \cite{Nak-etal:2018:APJ:}. 

The orthonormal components of the MF ${\bf B}=(B^{\widehat{r}},B^{\widehat{\theta}},B^{\widehat{\phi}} )$, measured by zero-angular momentum observer (ZAMO) can be determined by \cite{Tur-et-al:2021:PhRvD:}
\beq
B^{\widehat{i}} = \frac{1}{2} \eta_{ijk} \sqrt{g^{jj} g^{kk}} F_{jk},
\eeq
where $\eta_{ijk}$ is the Levi-Civita tensor in three dimensional space, and $F_{i j} = A_{j, i} - A_{i, j}$ is the antisymmetric tensor of the EM field. The non-vanishing orthonormal components of the parabolic MF take the form
\bea
B^{\widehat{r}} &=& \frac{B}{2} ~ \frac{\cos\theta}{|\cos\theta|} ~ r^{w-2},\\
B^{\widehat{\theta}} &=& - \frac{B}{2} \left(\frac{1 - |\cos\theta| }{\sin\theta} \right) \sqrt{f(r)} ~ r^{w-2} ~ w. 
\eea
The magnitude of MF $\textbf{B}$ as a function of $r$ and $\theta$ takes the form
\bea\non
|\textbf{B}| &=& \frac{1}{2} [B^2 r^{2w - 5}\, \{r - \frac{1}{2}  \left(4~ |\cos\theta| - 3 - \cos2\theta \right) \csc^2\theta \\\label{magnitude} &\times&  (r-2)\, w^2 \}]^{1/2}.
\eea
The realistic accretion states from GRMHD simulations that get magnetically saturated do form such a split monopole state that lives transiently until accretion destroys it again (i.e. a thin current sheet at the base of a paraboloidal jet that looks like a split monopole nearby the BH) \cite{Ripperda-etal:2022:ApJ:}. 
The realistic or simulated BH magnetospheres \cite{Tch-Nar-McK:2010:APJ:,Cri-etal:2020:AA:}  have a more complicated structure.  Despite this, our simplified model (\ref{mag_field}) captures their most  important features. 
One can distinguish three different regions around an accreting BH (see Fig.~\ref{fig_motiv}): the {\bf accretion disk/torus}, where the matter density is high $\rho\sim\rho_{\rm max}$ and matter dominates over the turbulent MF; {\bf corona region}, where the matter density is much lower $\rho\leq\rho_{\rm max}$, but the matter still dominates over the still turbulent MF; {\bf jet funnel}, where the matter component is very weak $\rho\leq10^{-6}\rho_{\rm max}$ and regular MF with parabolic shape dominates the region. The MF lines arising from the BH event horizon in equatorial plane are giving the boundary between the corona and jet funnel region. Corona-jet boundary condition is independent of magnetic intensity parameter $B$ and reads 
\beq
2^\xm - r^\xm \left( 1 - |\cos\theta| \right) = 0. \label{JetCorona}
\eeq
In our simplified Schwarzschild BH magnetosphere model (\ref{mag_field}), the MF lines are not rotating and light cylinder will not be formed. Mostly due to the density differences, the mean free path for charged particles differs dramatically in all three (disk/corona/jet) regions. It is interesting to note  that at least for M87 and Sgr~A* parameters, also the disk is collisionless due to the small mass accretion rate and low plasma mass density. In this article, we will assume the description of collisionless charged test particle dynamics can be well applied at least in jet funnel and corona regions and hence we will focus on these two regions leaving the description of accretion disk to a more advanced GRMHD \cite{Tch-Nar-McK:2010:APJ:,Nak-etal:2018:APJ:} or PIC \cite{Cri-etal:2020:AA:,Hir-etal:2021:APJ:} simulations.

\begin{figure*}
\includegraphics[width=\hsize]{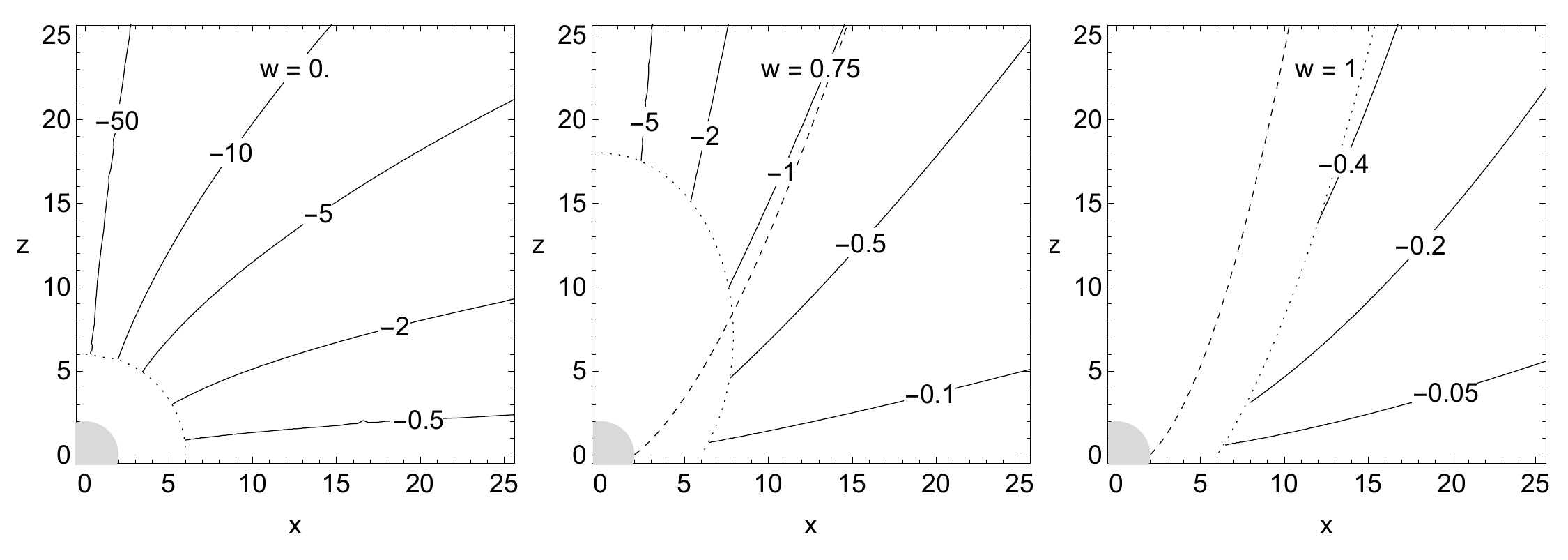}
\caption{For attracting LF ($\cb<0$), we plot the radial positions of stable circular orbits for charged particles in the off-equatorial plane. Different MF configurations $\xm=0,0.75,1$ have been plotted. The grey disk shows the BH horizon, the numbers on each solid curve indicate the values of the MF parameter $\cb$, while the points of these curves show the position of stable off-equatorial circular orbits. The dotted curve represents the offISCO for charged particles in the off-equatorial plane, while the dashed curve shows the corona/jet boundary.
\label{figISCOoff}
}
\end{figure*}

The charged test particle motion is described by the covariant Lorentz equation
\beq
	m \frac{D u^\mu}{\d \tau} = q F^{\mu}_{\nu} u^{\mu},
\eeq
where $u^{\mu}$ is the four-velocity of the particle with the mass $m$ and charge $q$, normalized by the condition $u^{\mu} u_{\mu} = -1$, $\tau$ is the proper time of the particle, and $F_{\mu \nu} = A_{\nu,\mu} - A_{\mu,\nu}$ is the antisymmetric tensor of the EM field. Using Hamiltonian formalism for the charged particle motion, we can write
\beq
  H =  \frac{1}{2} g^{\alpha\beta} (\cp_\alpha - q A_\alpha)(\cp_\beta - q A_\beta) + \frac{1}{2} \, m^2
  \label{particleHAM},
\eeq
where the kinematical four-momentum $\p^\mu = m u^\mu$ is related to the generalized (canonical) four-momentum $\cp^\mu$ by the relation
\beq
 \cp^\mu = \p^\mu + q A^\mu, \label{particleMOM}
\eeq
that satisfies the Hamilton equations in the form
\beq
 \frac{\d \x^\mu}{\d \af} \equiv \p^\mu = \frac{\partial H}{\partial \cp_\mu}, \quad
 \frac{\d \cp_\mu}{\d \af} = - \frac{\partial H}{\partial \x^\mu}. \label{Ham_eq}
\eeq
The affine parameter $\af$ of the particle is related to its proper time $\tau$ by the relation $\af=\tau/m$.

Due to the symmetries of the Schwarzschild spacetime (\ref{SCHmetric}) and the parabolic MF (\ref{mag_field}), one can easily find the conserved quantities that are the energy and the axial angular momentum of the particle and can be expressed as
\bea
 E &=& - \cp_t = m f(r) \frac{\d t}{\d \tau}, \\
 L &=& \cp_\phi = m \, r^2 \sin^2\theta \, \frac{\d \phi}{\d \tau} + \frac{q B}{2} r^{\xm} (1- | \cos\theta | ). \label{angmom}
\eea
The dynamical equations for the charged particle motion in the Cartesian coordinates can be found by the
transformation equations
\beq
x = r \cos\phi\, \sin\theta, \quad y = r \sin\phi\, \sin\theta, \quad z = r \cos\theta.
\eeq
Introducing for convenience the specific parameters, energy $\ce$, axial angular momentum $\cl$, and magnetic parameter $\cb$, by the relations
\beq
\ce = \frac{E}{m}, \quad \cl = \frac{L}{m}, \quad \cb = \frac{q B}{2m}, \label{ELB}
\eeq
one can rewrite the Hamiltonian (\ref{particleHAM}) in the form
\beq
H = \frac{1}{2} f(r) \p_r^2 + \frac{1}{2r^2} \p_\theta^2  + \frac{1}{2} \frac{m^2}{f(r)} \left[ V_{\rm eff}(r,\theta) - \ce^2 \right], \label{HamHam}
\eeq
where $V_{\rm eff}(r,\theta; \cl,\cb)$ denotes the effective potential given by the relation
\beq
V_{\rm eff} (r,\theta) \equiv f(r) \left[ 1 +\left(
 \frac{\cl}{r \sin{\theta} }
 - \cb\, r^\xm \, \frac{1- | \cos\theta|}{r \sin{\theta} }
 \right)^2\right]. \label{VeffCharged}
\eeq
The terms in the parentheses correspond to the central force potential given by the specific angular momentum $\cl$, and EM potential energy given by the magnetic parameter $\cb$. The Hamiltonian (\ref{HamHam}) can be divided into dynamical part $H_{\rm D}$ (first two terms containing  dynamical momenta $p_r,p_\theta$) and potential part $H_{\rm P}$ (last term).

The effective potential (\ref{VeffCharged}) shows clear symmetry $(\cl,\cb)\leftrightarrow(-\cl,-\cb)$ that allows to distinguish the following two situations
\begin{itemize}
\item[$-$] {\it minus configuration}, here $\cl>0, \cb<0$  (equivalent to $\cl<0, \cb>0$) -- MF and angular momentum parameters have opposite signs and the LF is attracting the charged particle to the $z$-axis, towards the BH.
\item[+] {\it plus configuration}, here $\cl>0, \cb>0$ (equivalent to $\cl<0, \cb<0$) -- MF and angular momentum parameters have the same signs and the LF is repulsive, acting outward the BH.
\end{itemize}
In this article, without loss of generality, we use the positive angular momentum of a particle $\cl>0$, while the magnetic parameter $\cb$ can be both positive or negative.


\begin{figure}
\centering
\includegraphics[width=0.75\hsize]{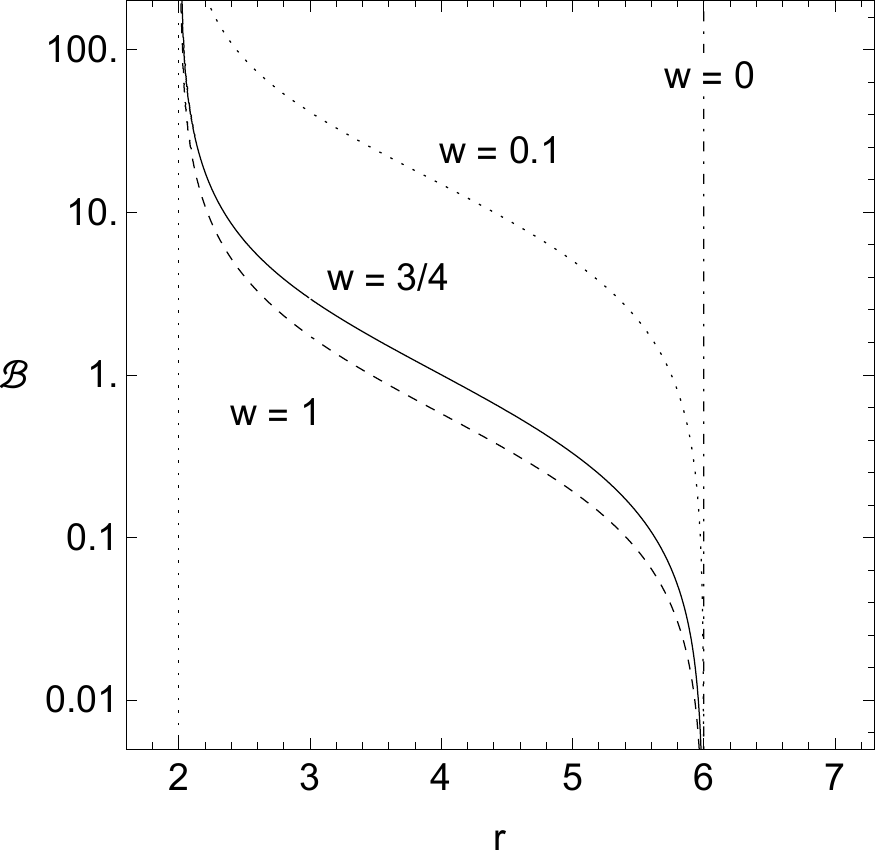}
\caption{
For repulsive LF ($\cb>0$), we plot the radial positions of the inISCOs for charged particles in equatorial plane, with different MF configurations $\xm=0,0.1,0.75,1$. The dotted vertical line shows the position of the BH horizon, while the dot-dashed vertical line indicates the position of the inISCO for a neutral particle. 
\label{figISCOin}}
\end{figure}

\begin{figure}
\centering
\includegraphics[width=0.75\hsize]{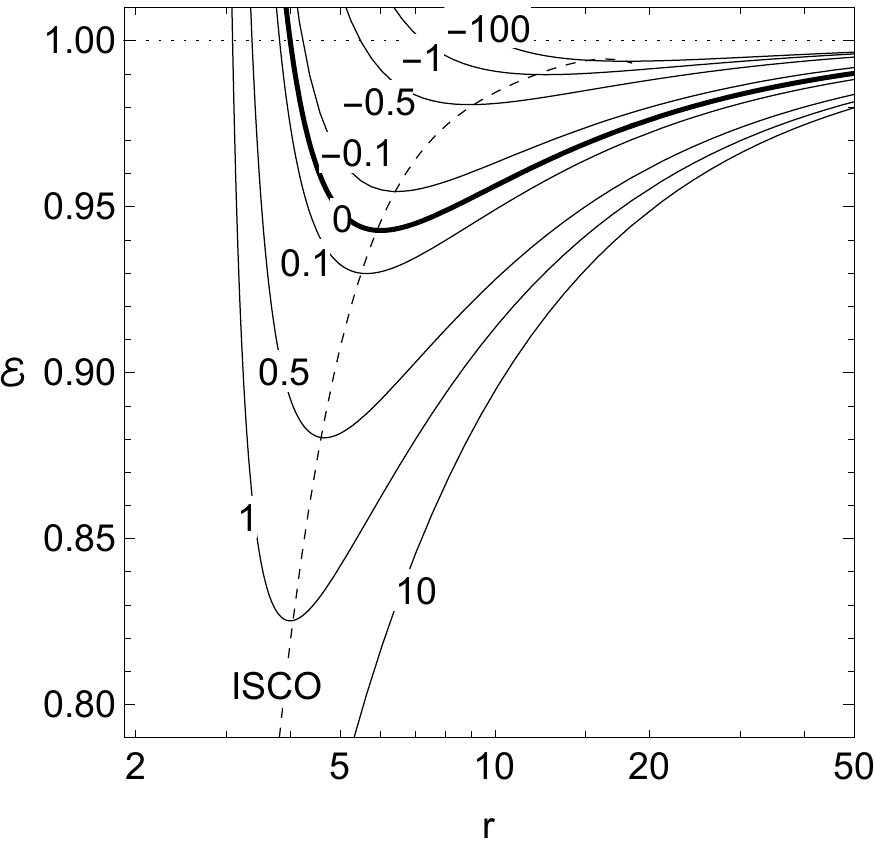}
\caption{Energy in effective potential minima, for off-equatorial ($\cb < 0$) and in-equatorial ($\cb > 0$) plane. We set the inclination parameter of the MF to $\xm=3/4$. The black thick curve is plotted for neutral particles. Each point of the dashed curve represents the position of offISCO (or inISCO), while the numbers on solid curves indicate the values of magnetic parameter $\cb$.
\label{figVeffMin}}
\end{figure}


\subsection{Off- and in-equatorial circular orbits}

The charged particle motion is confined by the energetic boundary given by
\beq
 \ce^2 = V_{\rm eff} (r,\theta; \cl,\cb). \label{MotLim}
\eeq
Let us properly investigate the features of the effective potential (\ref{VeffCharged}) represented in Fig. \ref{figVeff} that enables us to demonstrate the general properties of the charged particle dynamics, avoiding the necessity to solve the equations of motion. Since the $V_{\rm eff}(r,\theta)$ function is symmetrical above $\theta\in(0,\pi/2)$ and below $\theta\in(\pi/2,\pi)$ equatorial plane, we can focus on the region $\theta\in(0,\pi/2)$ only, and we can substitute $|\cos\theta|$ with $\cos\theta$ function here. The effective potential as a function of coordinates $x$-$z$ is plotted in Fig.~\ref{figVeff}.

The effective potential $V_{\rm eff}(r,\theta)$ function is zero at BH horizon
and limits are
\beq
 \lim_{x \to \infty} V_{\rm eff}(x,z) = 1, \quad
 \lim_{z \to \infty} V_{\rm eff}(x,z) = 1 + \frac{\cl^2}{x^2},
\eeq
for all values of magnetic parameters $\cb$ and $\xm \in(0,1)$. Only for value $\xm=1$, we have in the case of first limit $\lim_{x \to \infty} V_{\rm eff}(x,z) = 1 + \cb^2$.

The stationary points of the effective potential $V_{\rm eff}(r,\theta)$ function, where maxima or minima can exist, are given by the equations
\beq
  \partial_r V_{\rm eff}(r,\theta;\cl,\cb) = 0, \quad \partial_\theta V_{\rm eff}(r,\theta;\cl,\cb) = 0.  \label{extrem}
\eeq
These equations for the region above the equatorial plane $\theta\in(0,\pi/2)$ are
\bea
 (\cl + F) [ (\cl &+& F) + (r-2) (\xm - 1)F + \cl ] \csc^2\theta \nonumber \\
 &+& r^2 = 0, \\
 (\cl &+& F) (\cl \cos\theta -F)= 0,
\eea
where we define the supplementary function
\beq
 F (r,\theta) = \cb r^\xm (\cos\theta - 1).
\eeq
The term containing $|\cos\theta|$ in $V_{\rm eff}(r,\theta)$ function (\ref{VeffCharged}) will create the discontinuity in the first derivative of $V_{\rm eff}(r,\theta)$ with respect to $\theta$ in an equatorial plane $\theta=\pi/2$, and such points must be also added when we would like to examine $V_{\rm eff}(r,\theta)$ function extrema.

\begin{figure*}
\includegraphics[width=\hsize]{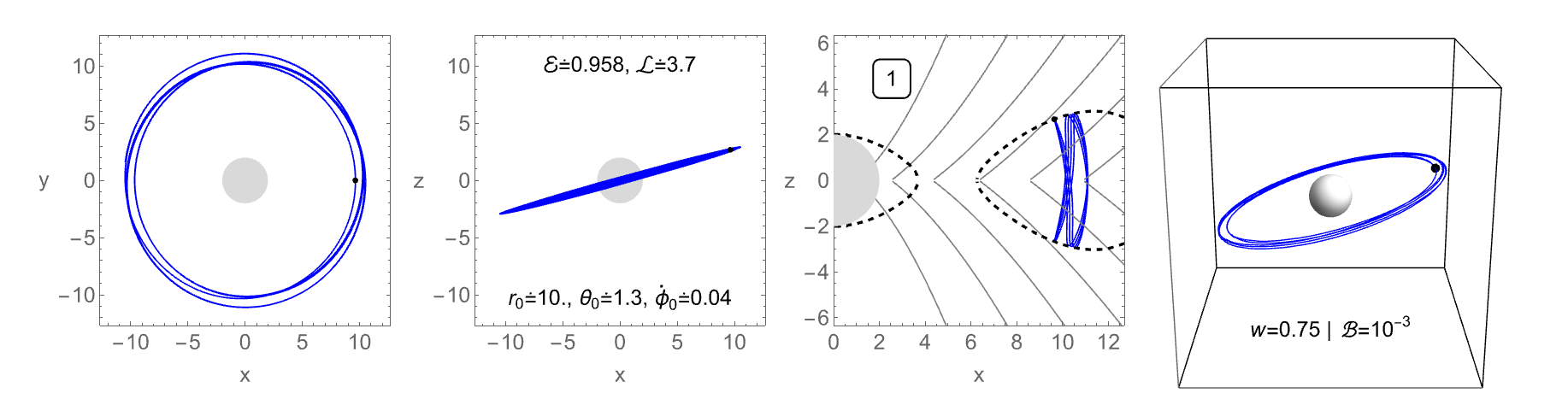}
\includegraphics[width=\hsize]{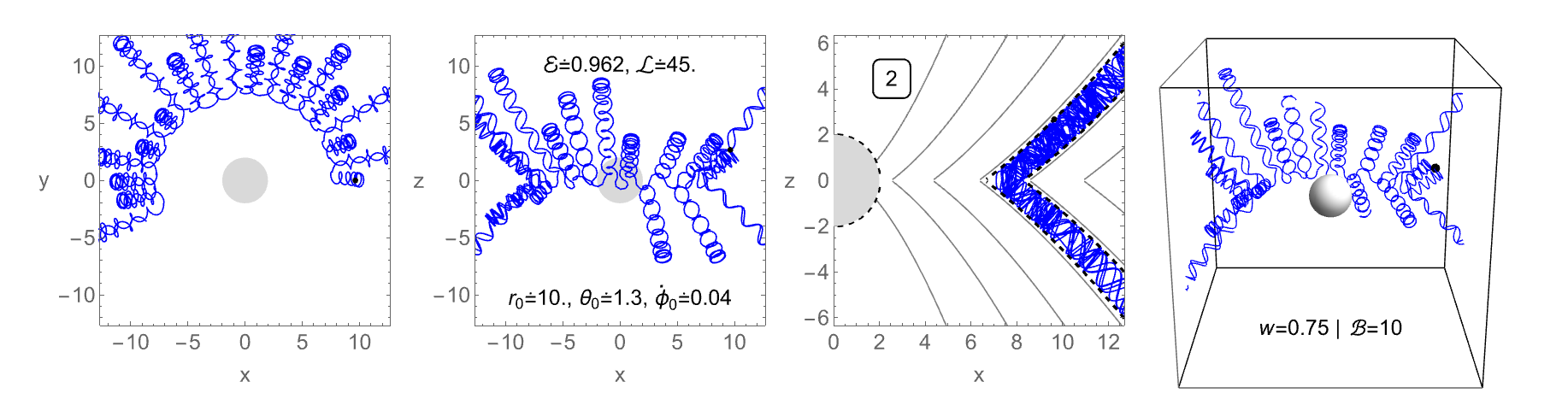}
\caption{Charged particles trajectories (blue) around BH with parabolic MF, representing two astrophysically relevant cases: when MF influence is negligible ($|{\cb}|{\ll}1$, first row), and when LF is dominating over gravity ($|{\cb}|{\gg}1$, second row). The gray circle represents the BH, gray curves MF lines, the dashed curve denotes the energy boundary, and the black dot the initial position of the particle. \label{figOrbitsREAL}}
\end{figure*}


There exist two classes of effective potential extrema solutions: off-equatorial plane extrema for attractive LF ($\cb<0$), and in-equatorial plane extrema for repulsive LF ($\cb>0$). 

For MF parameter $\cb<0$, we have minima located at off-equatorial plane 
\beq
\cos(\theta) = \frac{\cb r^\xm}{\cb r^\xm - \cl},
\eeq
while radial coordinate $r$ can be calculated using angular momentum condition $\cl=\cl_{\rm off}(r)$, where
\beq
\cl_{\rm off}(r) \equiv \frac{Y(r) + \sqrt{Y(r)^2 + 4 (r-3) r^2 }}{2 (r - 3) }.
\label{clofff}
\eeq
The new function $Y(r)$ is given by
\beq
 Y(r) = \cb\, r^\xm \left( \xm \left(2-r \right) + 2 \left(r -3 \right)\right).
\eeq
Off-equatorial plane minima are limited by the minimal distance from the BH -- so-called innermost stable circular orbits (offISCOs) given by the equation
\beq
\cb \cl \left(r \left(w(1-w) + 2 \right) + 2 (w-3) w \right) r^{w-1} - \cl^2 + 2 r = 0,
\eeq
where the particle angular momentum $\cl$ is given by Eq.~(\ref{clofff}). The radial positions of stable circular orbits as well as offISCO situated in the off-equatorial plane for different values of magnetic parameter $\cb<0$ and declination parameter $\xm$ are depicted in Fig.~\ref{figISCOoff}. Both offISCO and off-equatorial stable circular orbits get closer to the $z$-axis with the increase of the magnetic parameter $\cb$. 
The offISCOs are depending also on MF configuration, been close to the BH for pure magnetic monopole $\xm=0$ than in the case of parabolic MF $\xm=3/4$.

Existence of the off-equatorial circular orbits in the background of magnetized BHs with external asymptotically uniform or dipole MFs \cite{Kov-Stu-Kar:2008:CLAQG:,Kov-etal:2010:CLAQG:,Kov-Kop-Kar-Koj:CQG-2013} inspired the idea of investigation of the possible existence of charged off-equatorial toroidal structures orbiting around magnetized BH. One should note, that BHs can not have their own long-lived dipolar fields due to the no-hair theorem \cite{Lyu-Mck:2011:PhRvD:,Bransgrove-etal:2021:PhRvL:}. Radiation belts filled with energetic charged particles exists around many different astrophysical objects (mostly with dipole magnetosphere), so also in the case of BH one can search for observational consequences. As it can be seen in Fig.~\ref{figISCOoff}, the off-equatorial plane minima are located closer to the $z$-axis (and within the jet funnel) with increasing MF magnitude $\cb$ forming off-equatorial torodial structures in low density jet funnel filled with charged particle orbiting around BH on the off-equatorial circular orbits. 


For repulsive LF ($\cb>0$), we have minima located in equatorial plane $\theta =\pi/2$, while radial coordinate can be calculated using angular momentum condition $\cl=\cl_{\rm in}(r)$, where
\beq
\cl_{\rm in}(r) \equiv \frac{Y(r) + \sqrt{\cb^2 \xm^2 (r-2)^2 r^{2 \xm}+4 (r-3) r^2}}{2 (r-3)}. \label{clin}
\eeq
Effective potential minima in equatorial plane are limited by in-equatorial innermost stable circular orbit (inISCO), given by the equation
\bea
\cb^2 \xm^2 \left(r-2 \right) \left(\xm r^2-5 \xm r+6 \xm -r \right) r^{2 \xm} & \nonumber\\
-\BB \xm \left(\xm \left(r^2-5 r + 6 \right)-2 r^2 + 11 r - 18\right) r^\xm
\times & \nonumber\\
\times \sqrt{\BB^2 \xm^2 \left(r-2 \right)^2 r^{2 \xm} + 4 \left(r-3 \right) r^2} & \nonumber\\
   + 2 \left(r-6 \right) \left(r-3 \right) r^2  & = 0.
\eea
The radial position of inISCO for different magnitudes of the MF $\cb$ and various values of parameter $\xm$, situated in an equatorial plane are depicted in Fig.~\ref{figISCOin}. The inISCO radius is constant $r_{\rm inISCO}=6$ only in the case of split magnetic monopole magnetosphere $\xm=0$ \cite{Kol-Bar-Jur:2019:RAGtime:}.
For other values of declination parameter $\xm$ of MF, the radii of inISCOs lie in the interval $2\leq{r_{\rm inISCO}}{\leq}6$. For large MF magnitude, the inISCO radii to the BH horizon $r_{\rm inISCO}=2$, however, when the MF is negligible, we obtain the limiting case $r_{\rm inISCO}=6$. Similar behaviour can be observed for uniform MF \cite{Kol-Stu-Tur:2015:CLAQG:, San-Pan-San:2022:PhRvD:}. 

\begin{figure*}
\includegraphics[width=\hsize]{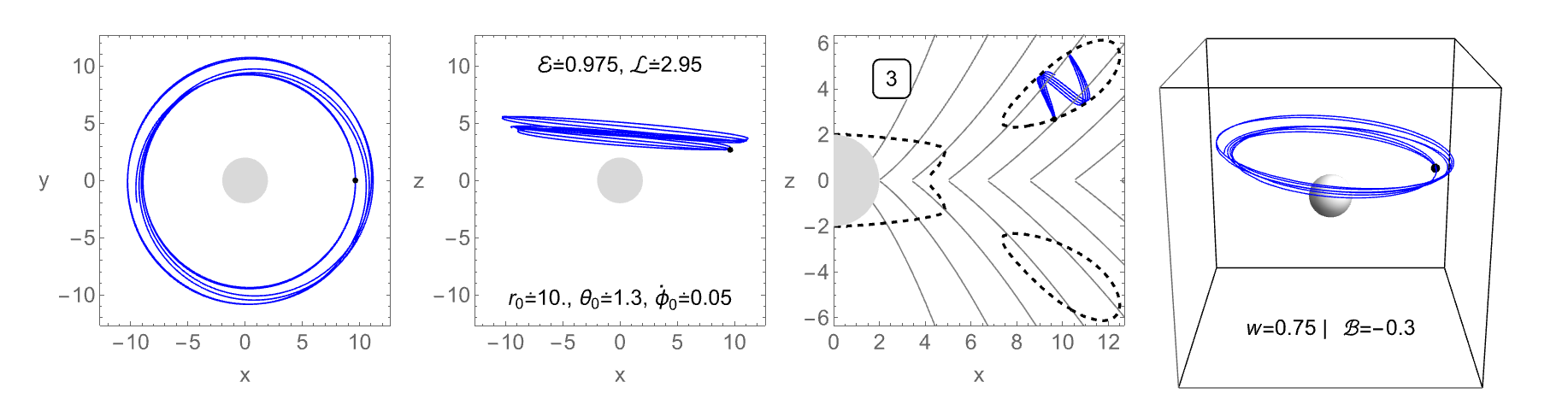}
\includegraphics[width=\hsize]{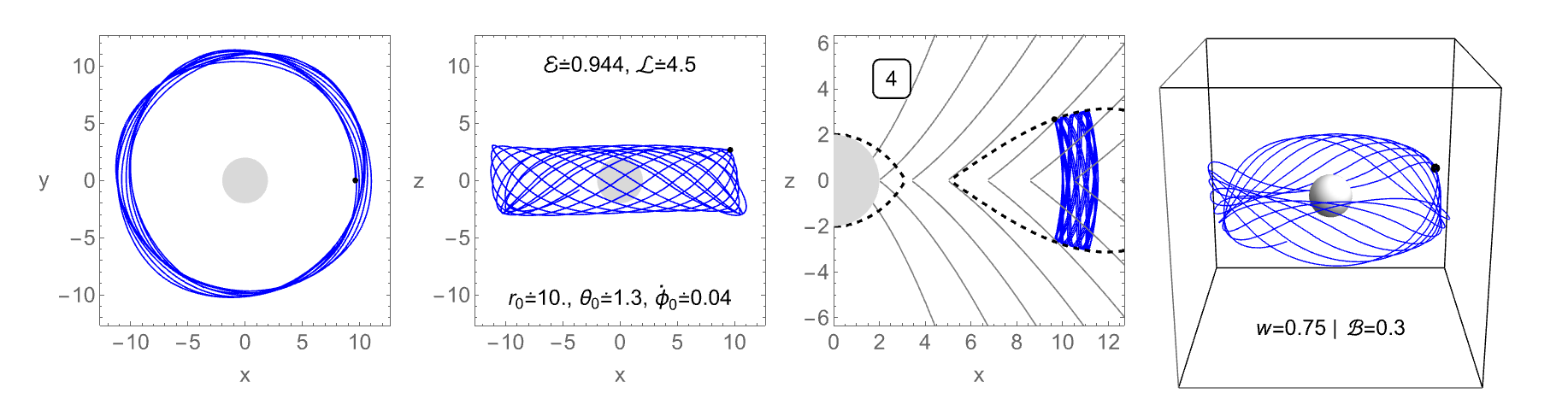}
\caption{Charged particle trajectories (blue) around BH with parabolic magnetosphere on orbits around the off- and in-equatorial plane minima. Gray disk represents the BH, gray curves the MF lines, the dashed curve the energy boundary and the black dot the initial position of the particle. \label{figOrbits}}
\end{figure*}


The graphical behaviour of the particle energy $\ce(r)$ as function of radial coordinate $r$ for circular orbits located in both off- and in-equatorial planes is shown in Fig.~\ref{figVeffMin}. Before the offISCO (or inISCO) position, the energy of stable orbits decreases with the increase of radial coordinate, but after the offISCO (or inISCO) position, the energy increases monotonically as the radial distance increases. As it can be seen from Fig.~\ref{figVeffMin}, charged particle on the off-equatorial plane circular orbit has lower binding energy than particle on circular orbit in the equatorial plane. One can conclude that off-equatorial circular orbits are less stable against perturbation than the circular orbits located in equatorial plane.

Limiting cases, when the MF is negligible ($|\cb|{\ll}1$) or when the MF is dominating over gravity ($|\cb|{\gg}1$) are more astrophysically prevalent - examples of particle orbits for these two cases are plotted in Fig~\ref{figOrbitsREAL}. In order to see clearly the interplay between magnetic LF and gravitation BH attraction leading to strongly non-linear and chaotic dynamics, one should use magnetic parameter $|\cb|\sim{1}$. The effect of attractive ($\cb<0$) and repulsive ($\cb>0$) LF is demonstrated on particle trajectories around off- and in-equatorial minima in Fig~\ref{figOrbits}. Effective potential minima can naturally serve as traps for particles and one could expect higher particle concentrations there, but the situation in/close to the equatorial plane must be described by a more complex approach than our single charged particle dynamics, for example by GRMHD of PIC simulations since Keplerian accretion disk formed by conductive plasma is expected to orbit BH in equatorial plane.


\begin{figure*}
\includegraphics[width=\hsize]{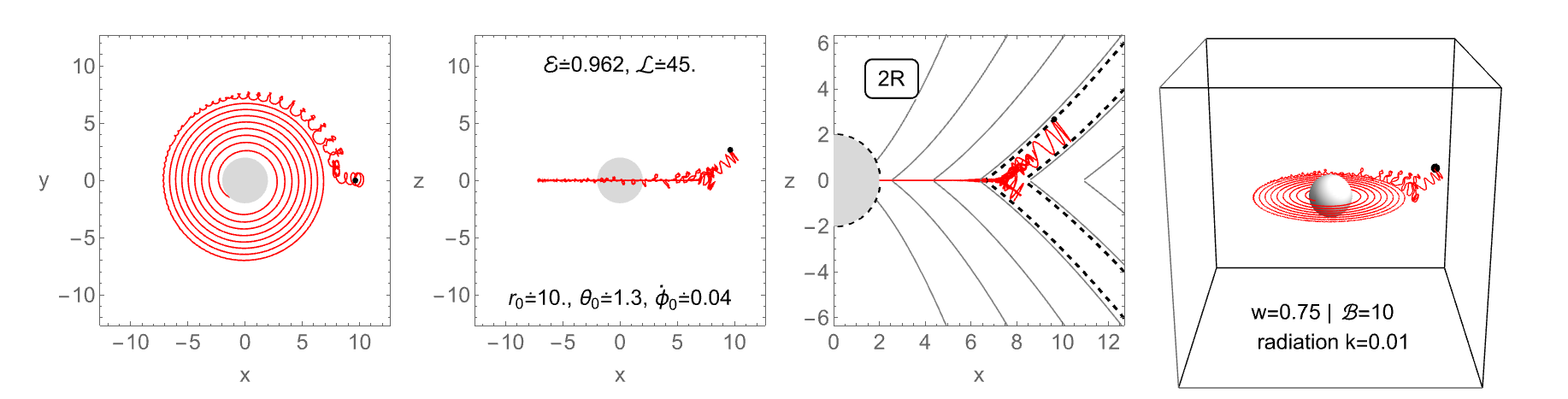}
\includegraphics[width=\hsize]{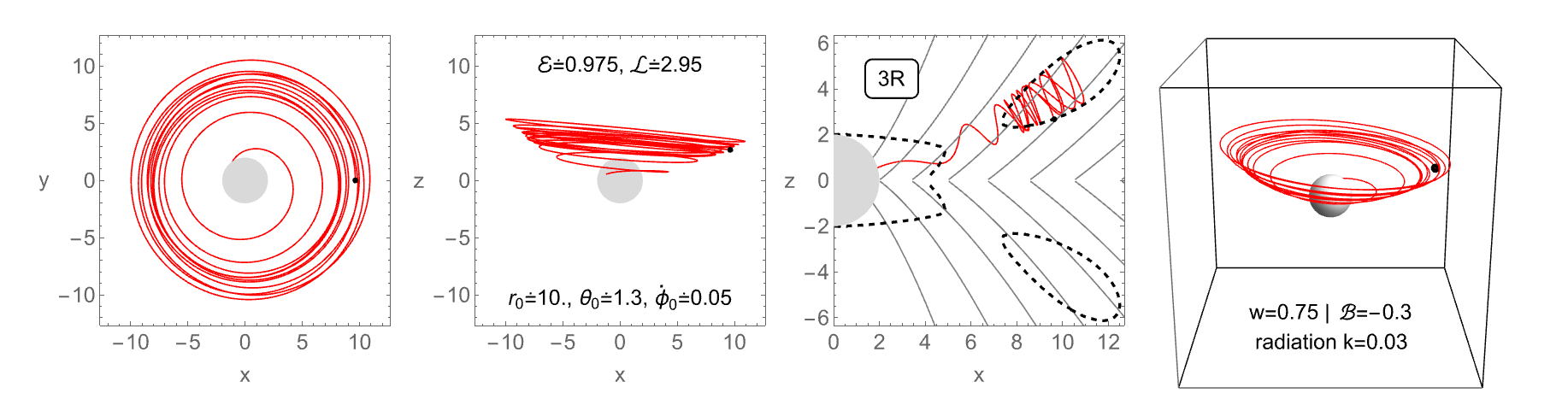}
\includegraphics[width=\hsize]{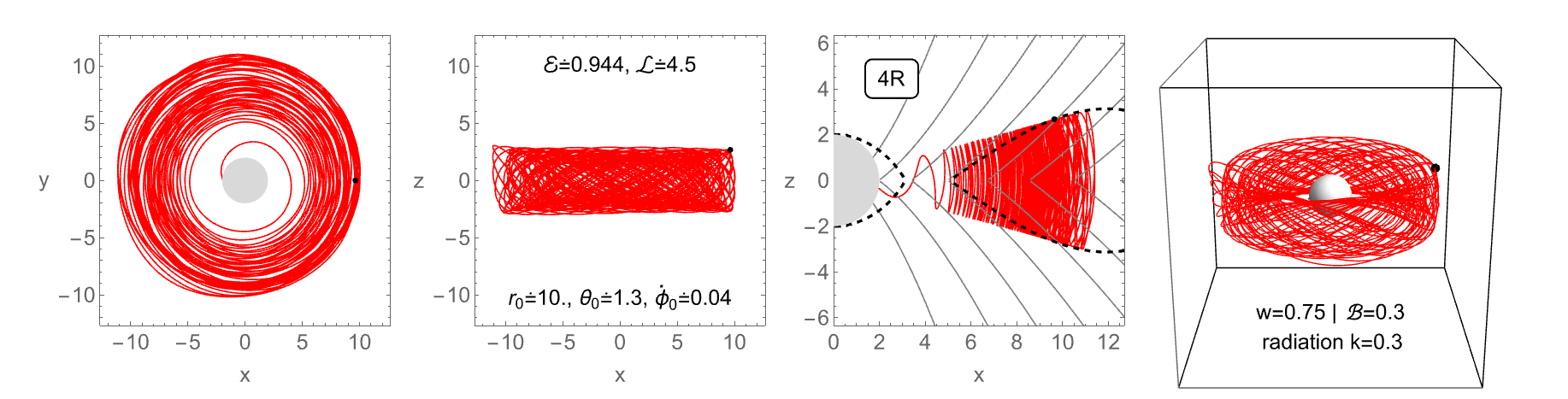}
\caption{Influence of RR on charged particle dynamics in parabolic ($\xm=0.75$) magnetosphere of  BH. The initial conditions for particle dynamics are the same as in Figs.~\ref{figOrbitsREAL} and \ref{figOrbits}. The radiation parameter $k$ has been chosen unrealistically large in order to clearly demonstrate the effect of RR force. Due to RR the particle spirals down to the BH. The damping of oscillations can be observed in the first row. Time scales related to RR  depend on MF parameter $\cb$ and specific charge, they are relevant for the strong MF, while negligible for weak one. \label{figRR}
}
\end{figure*}

\section{Radiating charged particle trajectories} \label{SecCHAOS}

\begin{figure*}
\includegraphics[width=0.9\hsize]{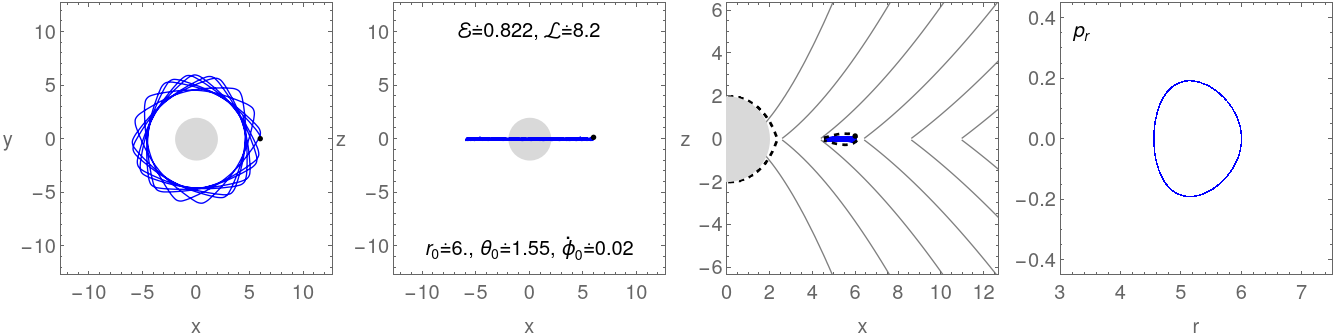}
\includegraphics[width=0.9\hsize]{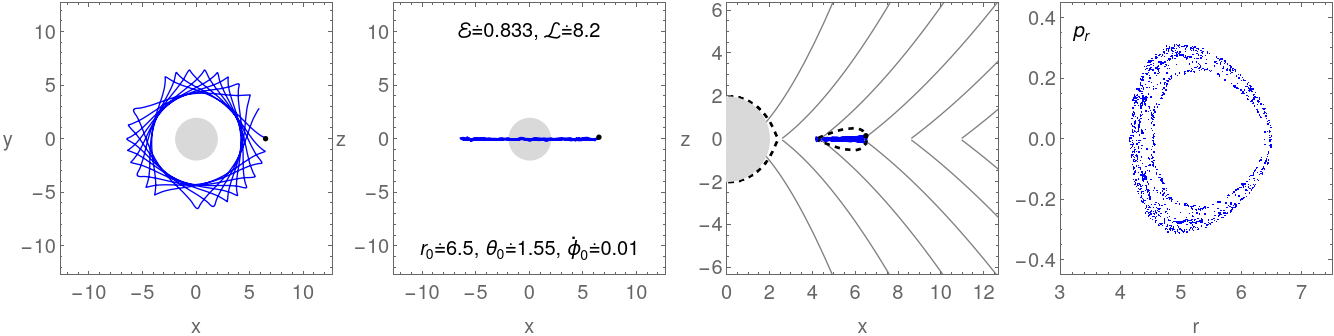}
\includegraphics[width=0.9\hsize]{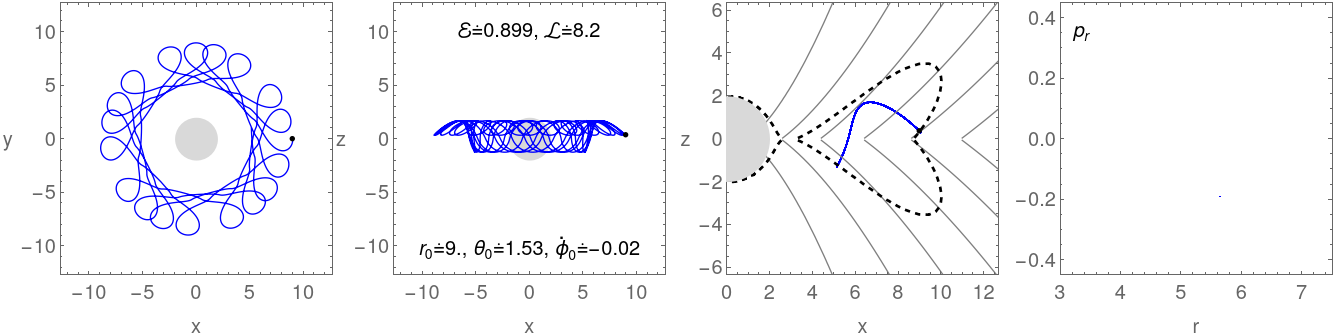}
\includegraphics[width=0.9\hsize]{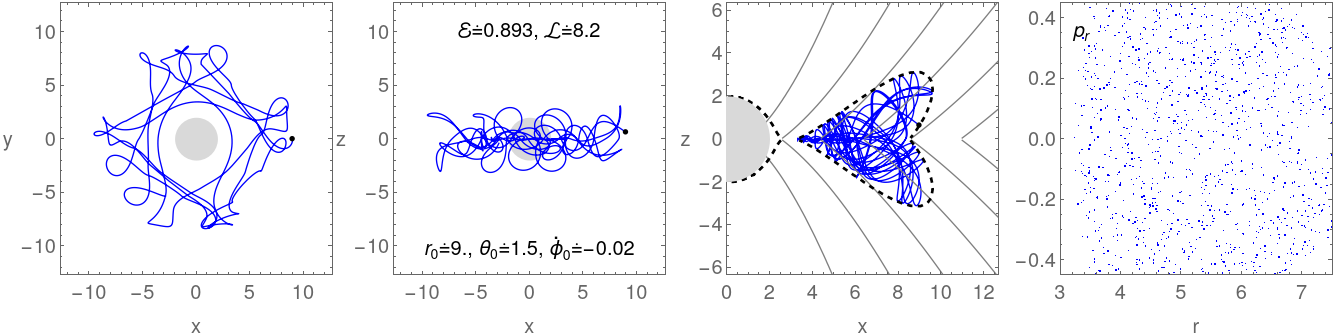}
\caption{
Charged particle chaotic dynamics (blue curve) in parabolic MF (gray parabolic lines) around BH (gray circle) can be well demonstrated by the PS plots (last column). In the first row, we see complete regular motion represented by smooth the Kolmogorov–Arnold–Moser (KAM) torus section. In the second row, we observe that this KAM torus is already starting to break and chaos is emerging with this orbit. 
In the third row some special (unstable) periodic orbit is presented, with only single point in PS. In the fourth row one can observe completely chaotic orbit, with points fully covering its phase space, limited only by particle's energy (dashed black curve). We use MF parameters $\cb=2$, $\xm=0.75$ for all orbits.
\label{figChaos}
}
\end{figure*}

\begin{figure*}
\includegraphics[width=\hsize]{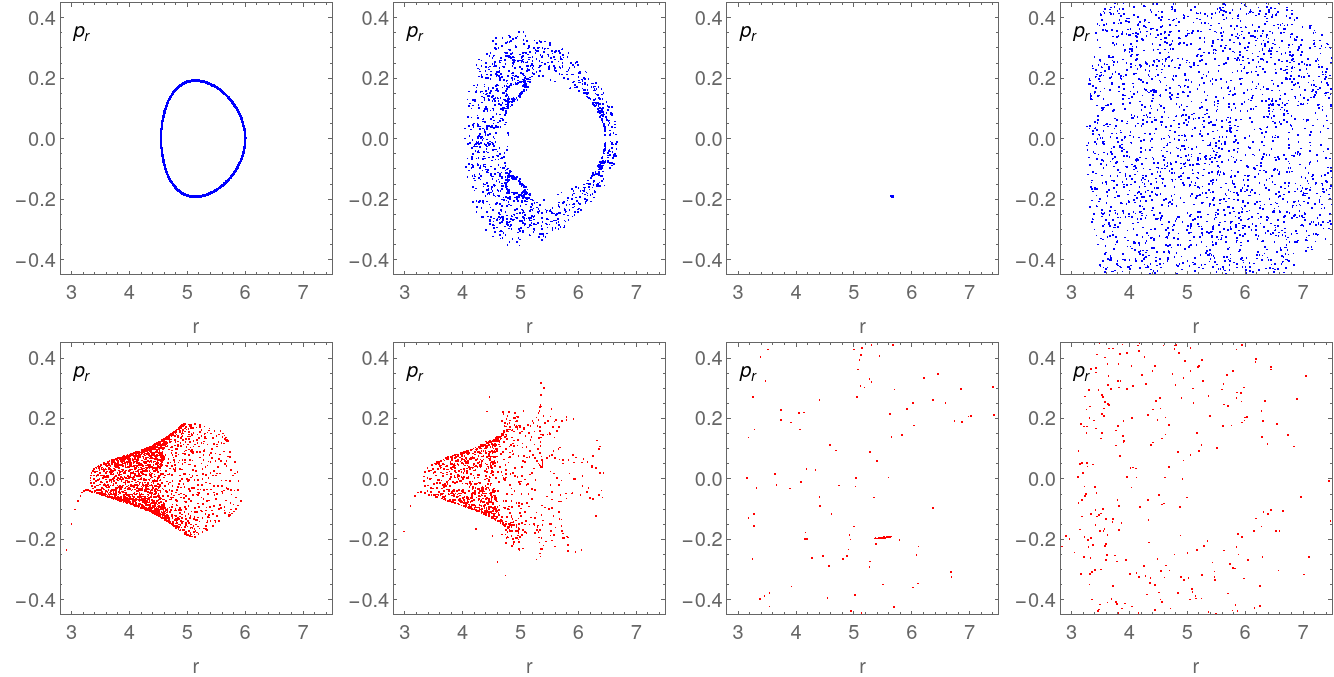}
\caption{
PSs for chaotic particle orbit without (blue first row) and with (red second row) the influence of RR. The same trajectories as in Fig.~\ref{figChaos} have been used. The RR is a non-linear damping force acting on a charged particle which can increase the chaoticity of a trajectory, but can also lead to the particle collapse to the BH. 
\label{figRRdamping} 
}
\end{figure*}

In many astrophysically relevant scenarios, one cannot neglect the effects of RR due to the synchrotron radiation of charges in the vicinity of BHs, which are believed to be immersed in an external MF. Equation of motion for a charged particle undergoing RR force in curved spacetime is non-trivial \cite{Poisson:2004:LRR:}. However, for elementary and subatomic particles, the equations can be simplified  to the form  \cite{Tur-etal:2018:APJ:}  
\bea \label{curradforce}
&& \frac{D u^\mu}{d \tau} = \frac{q}{m} F^{\mu}_{\,\,\,\nu} u^{\nu}  \\
&& + \,   \frac{q\,k}{m} \left(F^{\alpha}_{\,\,\,\beta ; \mu} u^\beta u^\mu + \frac{q}{m} \left( F^{\alpha}_{\,\,\,\beta}
F^{\beta}_{\,\,\,\mu} +  F_{\mu\nu} F^{\nu}_{\,\,\,\sigma} u^\sigma u^\alpha \right) u^\mu \right), \nonumber
\eea
where semicolon denotes the covariant coordinate derivative. The Eq. (\ref{curradforce}), which is a covariant form of the Landau-Lifshitz equation \cite{Lan-Lif:1975:CTF:}, is a habitual second order differential equation which satisfies the principle of inertia and does not contain runaway solutions. 
The exact equation of motion expressed for our particular parabolic magnetosphere are rather complex and hence we do not present exact formulas here, however we solve them numerically. The effect of RR is characterized by the radiation parameter
\beq
k=\frac{2}{3} \frac{q^2}{m}.
\eeq
Detailed discussion and derivation of (\ref{curradforce}) can be found in \cite{Tur-etal:2018:APJ:}. One should note, that in our approach the charged particle is radiating due to acceleration given by the LF which has many orders of magnitude stronger effect than the so-called geodesic synchrotron radiation \cite{Bre-etal:1973:PRD:,Tor-Dol:2022:PRD:}; in our approach such radiation on geodesic given by the tail integral \cite{Tur-etal:2018:APJ:} can be neglected. Detailed analysis of the particular case of the motion of charged particles around Schwarzschild and Kerr BHs immersed into an external asymptotically uniform MF was presented in \cite{Tur-etal:2018:APJ:,Kol-Tur-Stu:2021:PRD:}.

Trajectories of radiating charged particles around Schwarzschild BH immersed in parabolic MF are depicted in Fig.~\ref{figRR}. For both attractive ($\cb<0$) and repulsive ($\cb>0$) LFs, the charged particle will lose energy and angular momenta due to the RR force, which will force it to spiral down to the BH. For radiating particle trajectories plotted in Fig.~\ref{figRR}, we use the same initial conditions as for non radiating ones presented in Figs.~\ref{figOrbitsREAL} and \ref{figOrbits}. For repulsive ($\cb>0$) LF, we can observe oscillation damping due to reaction leading to particle's temporary stabilization on circular orbit (``orbital parking"). This ``parking" effect can be observed only in repulsive $\cb>0$ case and similar effect for repulsive LF has been already observed in the case of BH in uniform MF  \cite{Tur-etal:2018:APJ:}. For parabolic MF, the situation is different than for uniform one, here the circular parking orbit is not stable and particle will be slowly in-spiraling down to the BH. The RR force acts as a damping force - the oscillatory character of the motion of charged particles changes because of the loss of energy and angular momentum and radiating charged particles will always be captured by the BH. The off-equatorial plane minima are unstable for radiating particles and charged particles can stay oscillating around off-equatorial plane minima only for a limited amount of time, which could still be quite long since RR is relevant only for relatively strong MFs.


It is well known that the neutral test particle motion in Schwarzschild background is regular and fully integrable. On the other hand, the charged particle dynamics around magnetized BHs is generally chaotic \cite{Kov-Kop-Kar-Koj:CQG-2013,Pan-Kol-Stu:2019:EPJC:}. The motion of charged particles is regular (integrable) only for magnetic monopole field configuration $\xm=1$ due to monopole's spherical symmetry \cite{Kol-Bar-Jur:2019:RAGtime:}. Strongly non-linear and chaotic trajectories can be seen in the case of parabolic MF $\xm=3/4$, see lower row in Fig.~\ref{figOrbitsREAL}. {Still, in parabolic MF not all orbits are chaotic: the trajectories corresponding to minima of effective potential or with small epicyclic oscillations around circular orbits are regular \cite{Kol-Stu-Tur:2015:CLAQG:,Tur-Stu-Kol:2016:PHYSR4:}. }
Such almost circular orbits are very important from the astrophysical point of view as they govern the thin (Keplerian) accretion disks and the toroidal fluid configurations.

Charged particle chaotic behaviour can be analysed with the help of the Poincar\'{e} section (PS), presented in Fig.~\ref{figChaos}. From the distribution of the plotted points in the PS, one can distinguish whether  the motion is chaotic or not. For regular motion, the plotted points form a closed curve in the 2-dimensional $r-p_{r}$ plane, while in the case of the chaotic motion, PS points are distributed apparently randomly in the allowed region. When the energy of the particle is small, tori of small harmonic oscillation are not broken, see first row of Fig.~\ref{figChaos}. Such tori get broken when the energy of the particle is increased, see second row of Fig.~\ref{figChaos}, and the chaotic motion appears. The sea of chaos spreads in the PS with the increase of the particle energy, see fourth row of Fig.~\ref{figChaos}. The process of chaos increase with energy is sometimes interrupted by regular orbits, see example of unstable periodic orbit in third row of Fig.~\ref{figChaos}.

The comparison between PS for orbits of particles with and without the RR force is plotted in Fig.~\ref{figRRdamping}. { The RR force acts as a damping force leading the radiating particle towards the BH attractor, hence  trajectories at PS points are moving radially downwards in time, as one can see the visible accumulation of points eventually leading towards $r=2, p_r=-1$.} The radiation losses are highly non-linear during inspiral towards BH and particle may experience multiple resonances \cite{Muk-Kop-Luk:2022:XXX:}, which is demonstrated as structures in radiating particle PS. Unstable periodic orbit, third column in Fig.~\ref{figRRdamping}, will get perturbed by RR and become chaotic. Already chaotic, fourth column in Fig.~\ref{figRRdamping}, will remain chaotic.


\begin{figure*}
\includegraphics[width=\hsize]{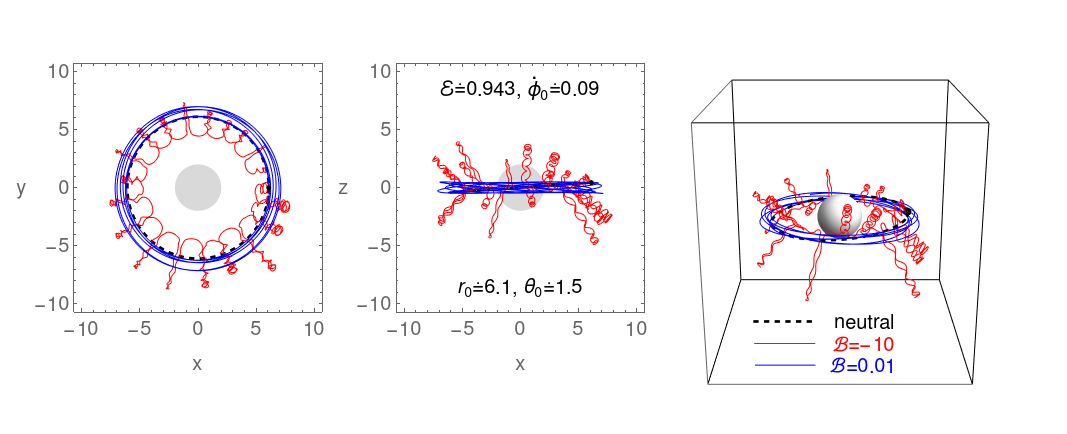}
\caption{
Example of ionization mechanism for neutral particles in Keplerian accretion disk. The LF and RR force acting on electrons and protons (ions) are significantly different. In the presented case, the proton (blue) is oscillating around effective potential minima in the equatorial plane and influenced by LF only slightly ($\cb=0.01$). 
On the other hand, the electron (red) is strongly influenced by the LF ($\cb=-10$) in the same MF, orbiting MF lines, while moving up, down, and around the BH. 
\label{figExampleION}
}
\end{figure*}

\section{Ionized Keplerian disks \label{secDisk}}

Single charged particle dynamics is a toy model for plasma description, but even with this simple tool one can test combined gravitational and EM effects of BH magnetosphere keeping control on numerical scheme, which could be clouded in more complex GRMHD or PIC numerical simulations. To demonstrate the influence of the MF on processes in BH vicinity, we consider an ionization of Keplerian accretion disk consisting of electrically neutral test particles following circular geodesics and orbiting a \Schw{} BH immersed in a parabolic MF, assuming the symmetry of Keplerian disk plane almost coinciding with the equatorial plane of the Schwarzschild geometry. 

Realistic ionization model of the originally neutral matter has been applied e.g. for the description of the magnetic Penrose process \cite{Par-Wag-Dad:APJ:1986:, Tur-etal:2020:APJ:}, where the original neutral particle 1 splits into two oppositely charged particles -- 2nd and 3rd, with charges $q_2$ and $q_3$, respectively. We thus assume conservation of the electric charge $0=q_2+q_3$, and~the canonical momentum
\beq
 \cp_{\alpha(1)} = \cp_{\alpha(2)} + \cp_{\alpha(3)}, \label{MPP}
\eeq
where $\cp_{\alpha(i)}$, $i=\{1,2,3 \}$ denotes the canonical momentum of 1st, 2nd and 3rd particles, respectively. Due to the charge conservation, the momentum conservation Eq. (\ref{MPP}) takes the form
\beq
 p_{\alpha(1)} = p_{\alpha(2)} + q_{2}A_{\alpha} + p_{\alpha(3)} + q_{3}A_{\alpha} = p_{\alpha(2)} + p_{\alpha(3)}. \label{MPP1}
\eeq
In many realistic scenarios one of the newly created charged particles is much more massive than the other one, $m_{2}/m_{3}\gg~1$, such as in the case of a neutral atom ionization.  Example of ionization mechanism for electrically neutral particle orbiting \Schw{} BH immersed in parabolic MF is shown in Fig.~\ref{figExampleION}. The magnitude of BH magnetosphere is set to $B$, but due to difference in masses of electrons and protons (ions), the LF ($\cb$ parameter) and RR forces ($k$ parameter) are very different. Therefore  more massive charged particle (proton or ion) takes almost all the initial momentum of the original neutral particle, and the dynamical influence of the lighter charged particle (electron) can be neglected, so that 
\beq
 p_{\alpha(1)}\approx p_{\alpha(2)}\gg p_{\alpha(3)}. \label{IonMech}
\eeq
In another realistic scenario, we can consider the Keplerian disk created by plasma modelled as a quasi-neutral soup of charged particles: electrons and ions, which are orbiting around the central object in circular orbits. If~the disk is dense enough, the~mean free path of charged particles is much shorter in comparison to the length of the orbit around the central object. This means that the charged particles orbit the BH together as a collective neutral body along circular geodesics as the influence of the MF is irrelevant to the motion, but~at the edges of the disk its density decreases substantially and the charged particle motion starts to be significantly influenced by the MF as the mean free path becomes comparable to the orbital length.

\begin{figure*}
\includegraphics[width=\hsize]{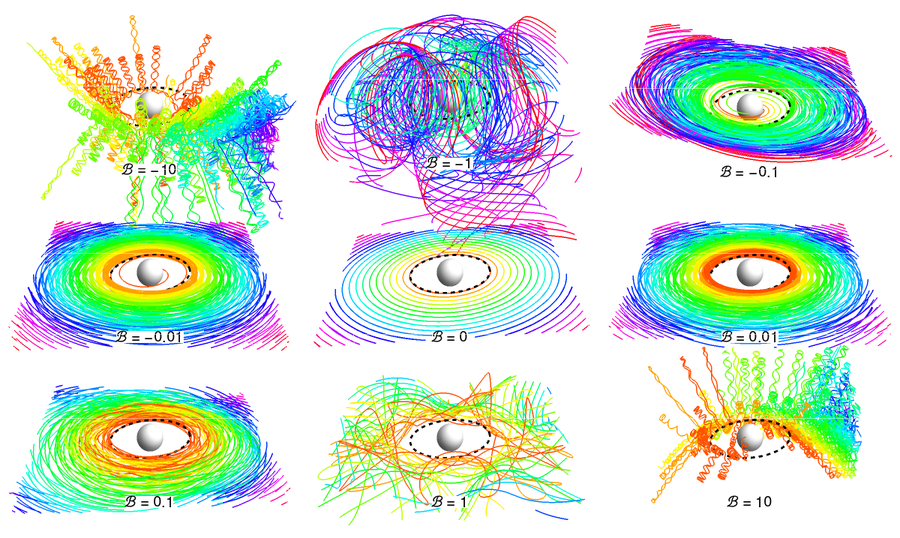}
\caption{
Evolution of thin Keplerian accretion disks around \Schw{} BH immersed into external  parabolic MF.  
Initially, the particles from accretion disk follow circular geodesics with slight inclination from the equatorial plane ($\theta_0=1.5$), while the MF is directed along the $z$-axis. We set the declination parameter of the parabolic MF lines to $\xm=3/4$, radiation parameter to $k=10^{-5}$, and the values for the MF parameter to $\cb=\{\pm10, \pm1, \pm 0.1, \pm 0.01, 0 \}$. The black dashed closed curves are the ISCOs for neutral test particles, indicating the inner edge of the Keplerian accretion disk. In the absence of the MF ($\cb=0$ case, middle figure), all the orbits remain in their circular shape and only the inclined razor-thin disk can be seen. 
If a slight EM interaction is switched-on ($\cb=\pm0.01$ cases, middle row), the charged particles creating a disk that is initially almost perpendicular to the MF lines start to oscillate with epicyclic frequencies around the circular orbit in both radial and vertical directions; the accretion disk becomes slightly thicker. Further, increasing the MF parameter ($\cb=\pm0.1,\pm1$ cases), the motion of charged particles becomes chaotic and the accretion disk is either destroyed or modified into a thick toroidal structure. For $\cb=-1$ case, all the particles are captured by the BH, and complete destruction of the Keplerian disk can be observed. The~LF dominates the particle motion when the parameter of the MF is large enough ($|\cb| \geq 10$ cases). The~charged particles spiral down and up along the MF lines, and move around the BH in the counter-clockwise ($\cb<0$) and clockwise ($\cb>0$) directions.
\label{figION}
}
\end{figure*}

\begin{figure*}
\includegraphics[width=\hsize]{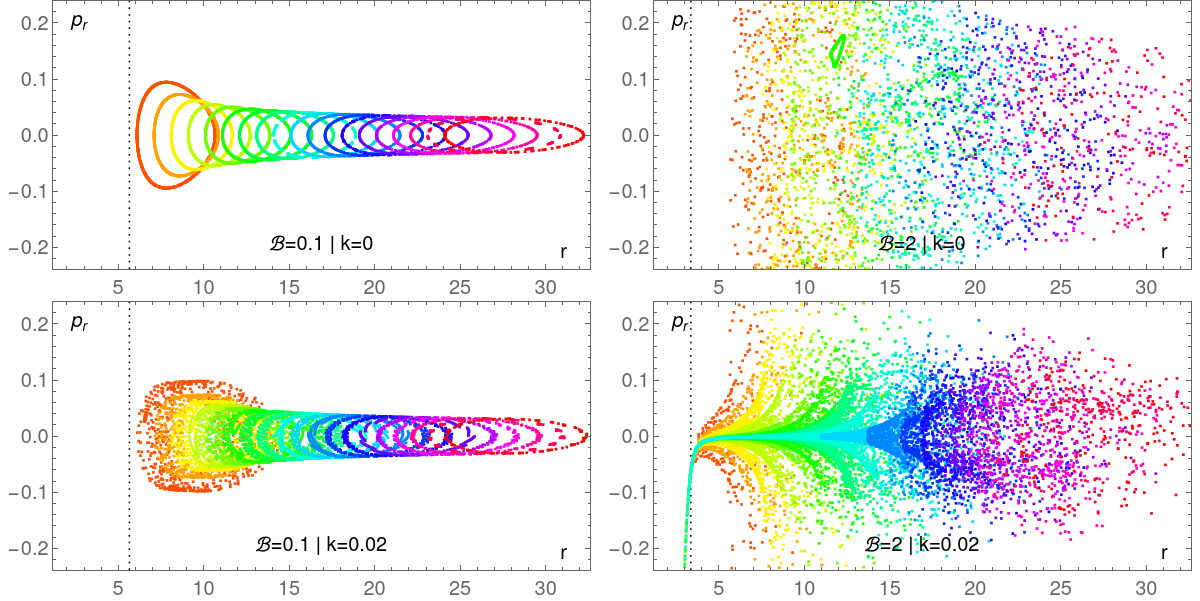}
\caption{
PSs for particles from ionized Keplerian disk, see Fig.~\ref{figION}. Two upper plots correspond to trajectories without RR: the left one ($\cb=0.1$) consists from regular trajectories forming closed loops, while the right one ($\cb=2$) from chaotic trajectories, where the points are distributed apparently randomly; the grey dotted vertical lines show the inISCO positions. In the two lower plots, the RR is included and we can see that some trajectories are attracted to the BH. All radiating particle trajectories may eventually end up in BH, however for small $\cb$ and $k$, or if the orbit is far away from the BH, the accretion takes much longer time. 
\label{figIONradiation}
}
\end{figure*}


In both scenarios neutral particles are ionized, while their mechanical momenta are  conserved. This simple ionization model (\ref{IonMech}) has been studied previously in the field of rotating Kerr BH with uniform MF \cite{Stu-Kol:2016:EPJC:,Kop-Kar:2018:APJ:,Tur-etal:2021:PRD:}. Chaoticity of the motion of the ionized matter orbiting Schwarzschild BHs has been studied in detail in~\cite{Pan-Kol-Stu:2019:EPJC:}.

Parabolic MF around non-rotating BH is given by $A_\phi$ component of the EM four-potential, hence particle's energy ($p_t$) is conserved during the ionization process (\ref{MPP1}). Particles from neutral Keplerian disk are on bounded orbits ($-p_t=E\leq1$), hence they can not escape to infinity after ionization since this would require ($E>0$), see (\ref{VeffCharged}). Charged particles are able to escape along open parabolic MF lines to infinity if they obtain initial kick in vertical direction -- such a scenario has been described in the uniform MF case in \cite{Zah-Fro-Sch:2013:PRD:,Shi-Kim-Chi:2014:PRD:,Bau-Veg:2021:CQG:}, where the parameters of the initial kick in vertical direction were calculated. For Kerr BH, the situation is different since here BH rotation generates ``electric'' component $A^t$ even in the case of a static MF configuration. The presence of the $A^t$ component leads to the energy redistribution in the formula (\ref{MPP}), which can lead the particle to escape from the bounded Keplerian circular orbit of BH with relatively high energy due to the so-called magnetic Penrose process \cite{Stu-Kol:2016:EPJC:,2019Univ....5..125T,Tur-etal:2020:APJ:,Tur-Kol-Stu:Symmetry:2022:}.


A test particle moving on the inclined circular orbit has initial position $x^\alpha$ and four-velocity $u_\alpha$, which can be written in the form 
\bea
 x^\alpha&=&(t,r,\theta,\phi)=(0,r_0,\theta_0,0),\\
 u_\alpha&=&(u_t,u_r,u_\theta,u_\phi)=(\ce,0,0,\cl).
\eea
Specific axial angular momentum $\cl$ and specific energy $\ce$ for neutral test particles on the inclined circular orbits from Keplerian accretion disk are given by~\cite{Wald:1984:book:}
\beq
 \cl_{(\mJ)} = \frac{r_0 \sin\theta_0}{\sqrt{r_0-3}}, \quad \ce_{(\mJ)} = \frac{r_0-2}{\sqrt{r_0^2 -3r_0}}. \label{CandLinSCHW}
\eeq
Due to a simple ionization condition (\ref{MPP1}), using the definitions of the specific energy and specific axial angular momentum (\ref{CandLinSCHW}), one can write for the specific axial angular momentum $\cl$ and the specific energy $\ce$ of the ionized test particle
\beq
  \cl_{(\mD)} = \cl_{(\mJ)} + A_{\phi} (r_{0},\theta_{0}) , \quad \ce_{(\mD)} = \ce_{(\mJ)}.
\eeq
Only the specific angular momentum $\cl$ is changed during the ionization, while the particle specific energy $\ce$ remains constant. Charged particles from the ionized Keplerian disk can either be captured by BH,  oscillate around circular orbit, or move chaotically along parabolic MF lines.


The influence of the MF parameter $\cb$ on the fate of the Keplerian disk orbiting a Schwarzschild BH in a position corresponding to the symmetry plane of the parabolic MF is demonstrated in Fig.~\ref{figION}. When the disk remains neutral or the MF is missing ($\cb=0$ case), all the orbits remain circular -- original inclined razor thin disk survives. For small values of the magnetic parameter ($\cb=\pm 0.01$ cases), the charged particles will be unsettled from original circular orbits and ionized disk will start to oscillate in both radial and vertical directions; the ionized disk becomes slightly thick due to the epicyclic motion of the charged matter. When the MF is slightly increased ($\cb=\pm0.1$ cases), complete destruction of the inner part of the Keplerian disk occurs in the attractive $\cb=-0.1$ case, as all the orbits end at the BH. In the repulsive $\cb=0.1$ case, the regular epicyclic motion has a tendency to chaoticity and the ionized disk becomes thick. Further increasing the MF parameter ($\cb=\pm1$ cases) the charged particle motion enters a fully chaotic regime causing a substantial thickening of the disk. For large values of the MF parameter ($\cb=\pm10$ cases), the LF becomes the leading force of the particle motion. LF in both attractive and repulsive cases leads to similar results: the charged particles go up and down along MF lines  realizing epicyclic motion around BH in  clockwise ($\cb>0$) or counter-clockwise ($\cb<0$) directions. 

The influence of the RR on the ionized Keplerian accretion disks orbiting Schwarzschild BH with parabolic MF can be observed for realistic values of the radiation parameter $k$ only in a very long integration time; in Fig.~\ref{figIONradiation} we have used $k=0.02$ to make RR effect more visible. The RR force reduces the chaotic behaviour of the particle motion. 
Neutral disk ionization and subsequent radiation losses do not lead to effective filling of the off-equatorial plane minima ($\cb<0$ case) with charged particles -- off-equatorial minima have higher energy than equatorial one, and off-equatorial minima are not attractors for radiating charged particles. Different processes similar to Earth's MF trapping charged particles from the solar wind in the Van Allen radiative belts can also be considered to form off-equatorial plane structures. It is worth mentioning that that the physics of pair cascades around BHs depend heavily on RR and Compton scattering \cite{Levinson-Cerutti:2018:A&A:,Chen-Yuan:2020:ApJ:,Crinquand-etal:2020:PhRvL:}.

\begin{figure*}
\includegraphics[width=\hsize]{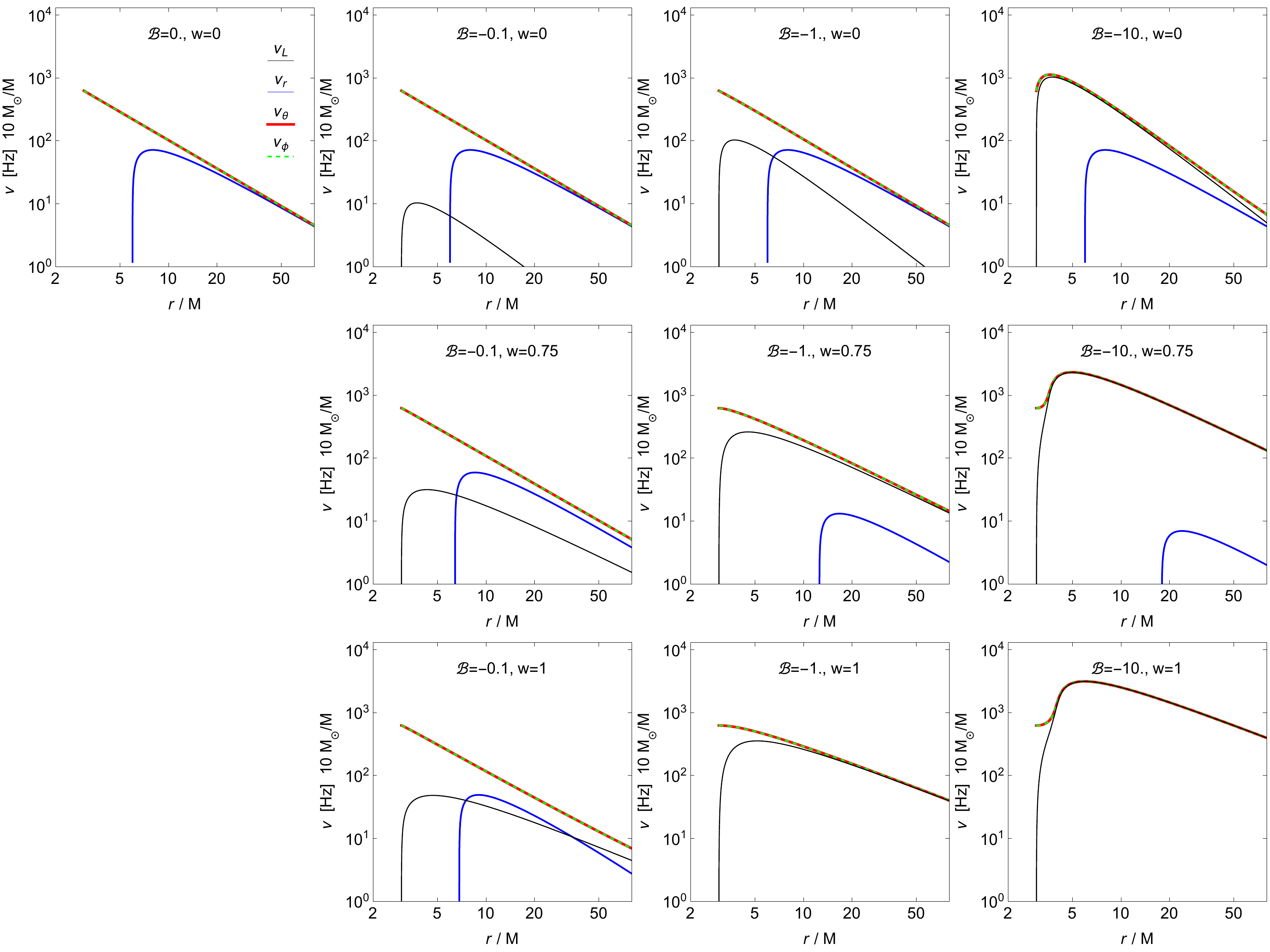}
\caption{
Off-equatorial radial profiles of the frequencies of small harmonic oscillations $\nu_r$, $\nu_\theta$, $\nu_\phi$, and $\nu_{L}$ of charged particle orbiting BH of the mass of $M=10 M_{\odot}$ in external parabolic MF for different values of the parameters $\cb$ and $w$.    
\label{OFFfigQPO}
}
\end{figure*}

\begin{figure*}
\includegraphics[width=\hsize]{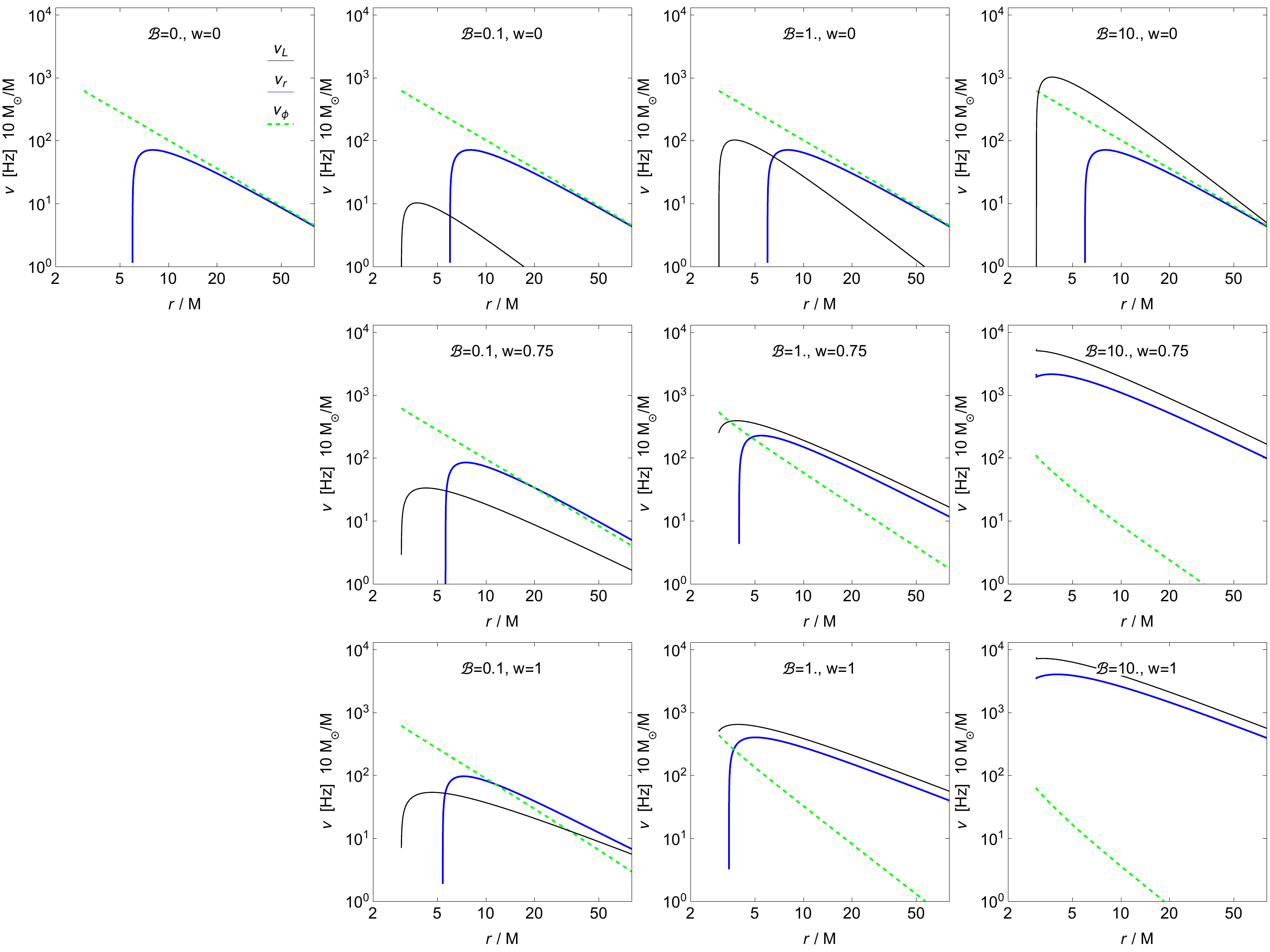}
\caption{In-equatorial radial profiles of the frequencies of small harmonic oscillations $\nu_r$, $\nu_\phi$, and $\nu_{L}$ of charged particle orbiting BH of the mass of $M=10 M_{\odot}$ in external parabolic MF for different values of the parameters $\cb$ and $w$. 
\label{INfigQPO}
}
\end{figure*}

\section{Particle oscillations and observed frequencies} \label{oscillations}

Ionization of particles from neutral accretion disk in the  equatorial plane may lead to the disk oscillations for $|\cb|<1$. If a test particle is slightly displaced from the equilibrium position, corresponding to a stable circular orbit, the particle starts to oscillate around the equilibrium position realizing thus epicyclic motion governed by the linear harmonic oscillations. In order to calculate the epicyclic frequencies of small perturbations to such orbits, we employ a linear stability analysis to compute the radial and vertical frequencies using a general method \cite{Kol-Stu-Tur:2015:CLAQG:}. Let us introduce a vector of canonical variables $y^{K} = (x^{i}, p_{i})$, so that we rewrite the equations of motion derived from the Hamiltonian (\ref{HamHam}) in the form
\beq
\dot{y}^{K} = f^{K}(y^N),\label{sol}
\eeq
where dot denotes the derivative with respect to the proper time $\tau$. We are interested in the solutions to the Eq. (\ref{sol}) near equilibrium position $y_{0}^{K}$ corresponding to a circular orbit, where $f^{K}(y_{0}^{N}) = 0$, except $N=3$. Linearizing $y^{K} = y_{0}^{K} + \zeta^{K}$, where $\zeta^{K}/y_{0}^{N}<<1$ denotes a small deviation vector, one gets a set of linear differential equations with constant coefficients
\beq
\dot{\zeta}^{K} = \left(\frac{\partial f^{K}}{\partial y^{N}} \bigg\rvert_{y=y_0} \right) \zeta^{K} + O(\zeta^{2}).\label{sol_A}
\eeq
We decompose the solutions to this system into eigenvalues and eigenvectors of the Jacobian matrix $\partial f^{K}/\partial y^{N}$. The eigenvalues $\lambda$ specify the rate at which trajectories with small differences in initial conditions separate. Thus, we evaluate Eq. (\ref{sol_A}) using the Hamiltonian (\ref{HamHam}) with the circular orbit values $y_{0}$ and find a pair of eigenvalues that we interpret as radial and vertical angular frequencies given by
\bea
 \omega^2_{\mir}  &=&  \frac{1}{2 g_{rr}} \left( \frac{\partial^2  H_{\rm p}}{\partial r^2} +  \frac{\partial^2  H_{\rm p}}{\partial \theta^2} - \sqrt{\chi} \right), \label{omegaR} \\
 \omega^2_{\mit}  &=&  \frac{1}{2 g_{\theta \theta}} \left( \frac{\partial^2  H_{\rm p}}{\partial r^2}  +  \frac{\partial^2  H_{\rm p}}{\partial \theta^2}  + \sqrt{\chi} \right), \label{omegaT}
\eea
where the unknown function $\chi$ takes the form
\beq
\chi = \left(\frac{\partial^2 H_{\rm p}}{\partial r^2} \right)^2 + \left(\frac{\partial^2 H_{\rm p}}{\partial \theta^2} \right)^2 
    + 4 \left( \frac{\partial^2 H_{\rm p}}{\partial r \partial \theta} \right)^2 - 2 \frac{\partial^2 H_{\rm p}}{\partial r^2} \frac{\partial^2 H_{\rm p}}{\partial \theta^2}.
\eeq
There exists the third fundamental angular frequency of the epicyclic particle motion, namely the Keplerian (axial) frequency $\omega_\mip$, given by
\beq
 \omega_{\mip} \equiv u^\phi = g^{\phi\phi} \left( \cl - q A_\phi \right). \label{omegaP}
\eeq

The radial and vertical angular frequencies (\ref{omegaR}-\ref{omegaT}) measured by a local observer for the case of in- and off- equatorial circular orbits have too complicated character when expressed explicitly. 
For $w=2$ and in equatorial plane, Eqs. (\ref{omegaR}) and (\ref{omegaP}) reduce to the uniform MF configuration \cite{Kol-Stu-Tur:2015:CLAQG:}, while $\cb=0$ leads to the neutral particle case \cite{Wald:1984:book:}. In contrast to the neutral case, there exists angular frequency so-called Larmor frequency for charged particles, which is associated with the MF only, and given by a relation
\beq
\omega_{\rm L} = \frac{q}{m} |\textbf{B}|,
\eeq
where $|\textbf{B}|$ can be found by Eq. (\ref{magnitude}). 
Due to the form of the Hamiltonian $H_{\rm P}$ (\ref{HamHam}) in parabolic MF, the first and second derivatives of $H_{\rm P}$ with respect to $\theta$ does not exist at the equatorial plane, hence the construction of $\omega_\mit$ is not possible at the equatorial plane. 
However, one can still perturb circular orbit in vertical directions around the equatorial plane, so that the perturbation will lead to the non-harmonic oscillations. Equations for the harmonic and non-harmonic oscillators are given, respectively, as follows  
\beq
 \ddot{\theta} + \omega_{\mit}^2 \, \theta = 0, \qquad  \ddot{\theta} + C_\theta \, {\rm sgn}(\theta) = 0, \label{aharmonic}
\eeq
where $C_\theta$ is a constant. 
Large vertical perturbation from equatorial plane lead to chaotic dynamics, see the second row in Fig.~\ref{figOrbitsREAL}.

\begin{figure*}
\includegraphics[width=0.9\hsize]{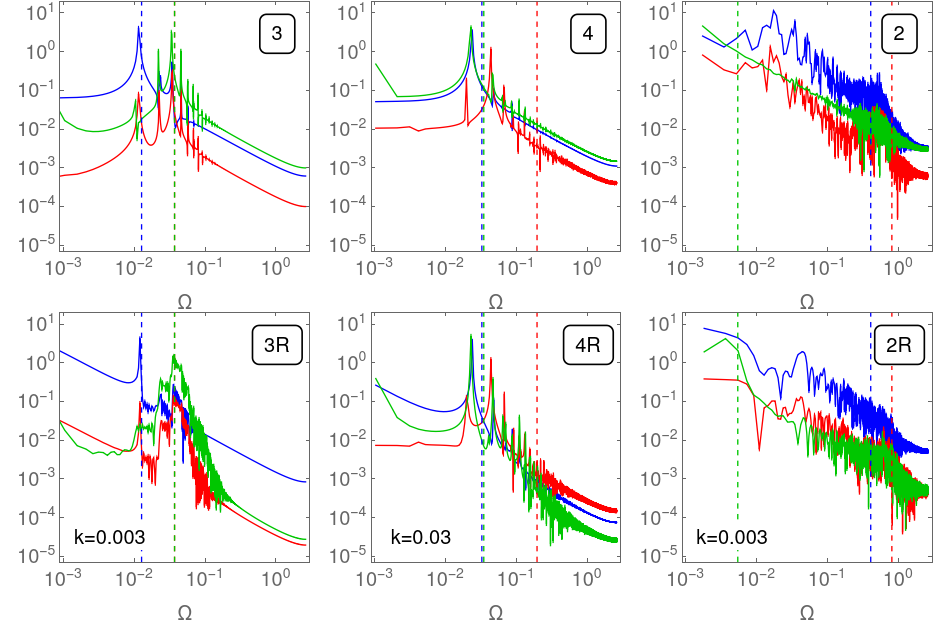}
\caption{
Power Spectral Density for  three $r(\tau)$ (blue), $\theta(\tau)$ (red) and $\phi(\tau)$ (green) coordinates along the particle's trajectory.  Fundamental frequencies calculated analytically for circular orbit and presented in Figs.~\ref{OFFfigQPO} and \ref{INfigQPO} are plotted as vertical dashed lines. The top row of figures corresponds to  trajectories without RR $k=0$, see, also Figs.~ 
\ref{figOrbitsREAL} and \ref{figOrbits}. Cases $3$ and $4$ represent regular trajectories around off- and in-equatorial plane minima with distinguishable main peaks; The case $2$ corresponds to the chaotic trajectory. Second row of figures is plotted for radiating particles, where the higher frequency peaks (overtones) are flattened due to radiation damping and the spectrum is more smooth (noisy) in higher frequencies. \label{FFtest}
}
\end{figure*}

The locally measured angular frequencies $\omega_\mir$, $\omega_\mit$, $\omega_\mip$ and $\omega_{\rm L}$ are connected to the angular frequencies measured by the static distant observers, $\Omega_{\beta}$, by the gravitational redshift transformation
\beq
 \Omega_{\beta} = \frac{\d \aaa_\beta}{\d t} = \omega_{\beta} \, \frac{\d \tau}{\d t} = \omega_{\beta} \, \frac{f(r)}{ \ce(r)},
\eeq
where $(f(r) / \ce(r) ) $ is the redshift coefficient, given by the  function $f(r)$ and the particle specific energy at the circular orbit $\ce(r)$. If the fundamental frequencies of the small harmonic oscillations related to the distant observers, $\Omega_{\beta}$, are expressed in the physical units, their dimensionless form has to be extended by the factor $c^3/GM$. Then the frequencies of the charged particle radial and latitudinal harmonic oscillations measured by a distant observer in physical units are given by
\beq
    \nu_{\beta} = \frac{1}{2\pi} \frac{c^3}{GM} \, \Omega_{\beta}.
\eeq
The behaviour of the fundamental frequencies $\omega_\mir, \omega_\mit, \omega_\mip$ and their ratios can help us to distinguish different shapes of charged particle epicyclic orbits in the vicinity of a stable circular orbit. 
 
 According to the Newtonian theory of gravitation, all the frequencies are equal, $\omega_\mir=\omega_\mit=\omega_\mip$, giving ellipse as the only possible bounded trajectory of a test particle around a gravitating spherically symmetric body. For uncharged particles moving around a \Schw{} BH, the relation $\omega_\mir<\omega_\mit=\omega_\mip$ holds, and there exists a periapsis shift for bounded elliptic-like trajectories implying the effect of relativistic precession that increases with decreasing radius of the orbit as the strong gravity region is entered \cite{Ste-Vie:1999:PHYSRL:}. For charged particle frequencies around BH in presence of MF, one gets in general $\omega_\mir\neq\omega_\mit\neq\omega_\mip$, see \cite{Kol-Stu-Tur:2015:CLAQG:,Tur-Stu-Kol:2016:PHYSR4:,Kol-Tur-Stu:2017:EPJC:}.

The radial $\nu_\mir$, vertical $\nu_\mit$, orbital $\nu_\mip$ and Larmor $\nu_{\rm L}$ frequencies measured by a distant observer in off-equatorial plane ($\cb<0$) for different MF configurations as a function of radial coordinate $r$ are plotted in Fig.~\ref{OFFfigQPO}. We see that orbital $\nu_\mip$ and vertical $\nu_\mit$ frequencies coincide for all considered MF configurations. When the attractive LF is weak, the Larmor frequency is smaller than the orbital frequency, however, it coincides with the orbital frequency as the magnetic parameter $\cb$ is increased. For parabolic MF ($\xm=3/4$), we can see that the radial frequency $\nu_\mir$ is changing dramatically with the parameter $\cb$, while the vertical $\nu_\mit$ and orbital $\nu_\mip$ frequencies remain almost the same. For strong attractive LF ($\cb<-1$), the charged particle  moves relatively slowly around BH at the off-equatorial plane minima, while oscillating relatively rapidly around MF lines. The radial frequency is too small for the case of paraboloidal MF with $\xm=1$, so the radial profiles start to appear only far away from the BH, which is in concord with the results presented for offISCO in Fig.~\ref{figISCOoff}.

Fundamental frequencies for the case of in-equatorial plane ($\cb>0$) are plotted in Fig.~\ref{INfigQPO}. When the repulsive LF is weak, the Larmor frequency $\nu_{\rm L}$ is the smallest among all the frequencies, while the radial $\nu_\mir$ and orbital $\nu_\mip$ frequencies coincide. However, when the repulsive LF is strong ($\cb>1$), the radial $\nu_\mir$ and Larmor $\nu_{\rm L}$ frequencies have  different profiles in strong gravity regime close to the BH, but the frequencies $\nu_\mir$ and $\nu_{\rm L}$ will come closer as the radial distance increases. 



The effect of RR on Fourier spectra (Power Spectral Density) calculated from charged particle trajectories is shown in Fig.~\ref{FFtest}. Particles moving close to circular orbits have main peaks at fundamental frequencies given by  Eqs.~(\ref{omegaR}) - (\ref{omegaP}). The higher harmonics (overtones) are also shown. The RR damping force does not change main frequency peaks, but it lowers the higher harmonics making the spectrum more flat and diluted in high frequency range (see cases $3$ and $4$ in Fig.~\ref{FFtest} in comparison with the cases $3R$ and $4R$, respectively). The RR makes the motion more irregular (chaotic) on short time scales. For both attractive ($\cb<0$ case 3) and repulsive ($\cb>0$ case 4) LFs, the orbital motion ($\phi$, green) is considerably suppressed in high frequency range. 
We interpret this effect as follows: since the azimuthal motion (in $\phi$ direction) in quasi-circular orbits has the highest curvature, hence larger acceleration with respect to other directions of the motion, this component of the motion radiates away faster. Due to RR the trajectory becomes more smooth on short time scales, since the high curvature modes are dumped. 
Similar effect of RR damping can also be seen for chaotic trajectory (case 2) due to the same reason. RR increases the chaoticity in already chaotic case only slightly, although  detectable using the tools from non-linear analysis \cite{Pan-Ada-Mar:2022:EPJST:}.


\section{Astrophysical estimates} \label{SecASTRO}

%
In order to relate our results to realistic astrophysical scenarios, in this section we provide estimates of the most relevant parameters of the discussed model in astrophysical situations. 
%
%
The MF strengths close to the BH surface can be estimated from the Eddington value (MF energy density coinciding with the density of accreting plasma) \cite{Rees:1984:ARAA:}
\beq
 B_{\rm Edd} \sim 10^4 \left( \frac{M}{10^9 M_{\odot}} \right)^{-1/2} {\rm Gs}.
\eeq
According to \cite{Daly:2019:APJ:,Pio-etal:2020:arXiv:}, the characteristic values of the MFs near the stellar mass and SMBHs are comparable with  the Eddington value. In Table~\ref{Tab1} we give the values of $B_{\rm Edd}$, orbital timescale $\tau_c$ of the particle at the radius $r=6M$ for two representative BH masses: stellar mass BH ${10}~M_{\odot} $ and SMBH ${10^9}~M_{\odot}$.

\begin{table}[!ht]
\begin{center}
\begin{tabular}{c c @{\qquad} c c } 
\hline
& BH mass & orbital time & $B_{\rm Edd}$  \\
\hline
stellar-mass BH & $10^1 M_{\odot}$ & $10^{-3}~{\rm{s}}$ & $10^{8}~{\rm{Gs}}$  \\
SMBH & $10^9 M_{\odot}$ & $10^{4}~{\rm{s}}$ & $10^{4}~{\rm{Gs}}$ \\
\hline
\end{tabular}
\caption{Typical size, inISCO orbital timescale, and Eddington MF for stellar and SMBHs, representing characteristic values of many BH candidates. \label{Tab1}} 
\end{center}
\end{table}

\begin{table}[!ht]
\begin{center}
\begin{tabular}{l l @{\quad} l @{\quad} l @{\quad} l @{\quad} l } 
\hline
MF $B$  & $\BB_{\rm electron}$ & $\BB_{\rm proton}$ & ~$\BB_{\rm Fe+}$~ & $\BB_{\rm dust}$ \\
\hline
 $ {10~{M}_{\odot}} $ & & & \\
\quad $B=10^8$~Gs & $10^{10}$ & $10^7$ & $10^5$ & $10^{-2}$ \\ 
\quad $B=10^4$~Gs & $10^6$ & $10^3$ & $10^1$ & $10^{-6}$ \\
\quad $B=10^0$~Gs & $10^2$ & $10^{-1}$ & $10^{-3}$ & $10^{-10}$ \\ 
\quad $B=10^{-5}$~Gs & $10^{-3}$ & $10^{-6}$ & $10^{-8}$ & $10^{-15}$ \\
\hline
$ {10^9~{M}_{\odot}} $ & & & \\
\quad $B=10^4$~Gs & $10^{14}$ & $10^{11}$ & $10^9$ & $10^{2}$ \\
\quad $B=10^0$~Gs & $10^{10}$ & $10^{7}$ & $10^{5}$ & $10^{-2}$ \\ 
\quad $B=10^{-5}$~Gs & $10^{5}$ & $10^{2}$ & $10^{0}$ & $10^{-7}$ \\
\hline
\multicolumn{2}{l}{Sgr~A* ($ {4 \times 10^6~{M}_{\odot}} $)} & & \\
\quad $B=10$~Gs & $10^{9}$ & $10^{6}$ & $10^4$ & $10^{-3}$ \\
\hline
\end{tabular}
\caption{
Magnetic parameter $\BB$, Eq. (\ref{BB-param}), for various MF magnitudes and different charged particles: electron, proton, ionized iron atom (one electron lost), dust grain (one electron lost, $m\sim 10^{-18}$~g). Stellar mass $M={10~{M}_{\odot}}$, SMBH  $M={10^9~{M}_{\odot}}$, and the Galactic centre BH (Sgr~A*) cases are considered.
\label{Tab2}
} 
\end{center}
\end{table}

The relative ratio between gravitational and magnetic LFs is represented by the parameter $\BB$. Restoring the world constants in Eq. (\ref{ELB}), the dimensionless parameter $\BB$, widely used in the present paper, takes the form
\beq 
\BB = \frac{|q| B G M}{2 m_e c^4}, \label{BB-param}
\eeq
reflecting the relative influence of the gravitational and MFs on the charged particle motion. 
As an example, let us estimate $\BB$ for an electron around Galactic centre SMBH Sgr~A*, which is currently one of the best known SMBH candidates. The equipartition value of the MF strength at the event horizon scale of Sgr~A* is usually estimated to be around $10-100$ Gs \cite{Johnson-etal:2015:Science:,Eatough-etal:2013:Natur:}. Mass of the BH is measured to be around $4 \times 10^6 M_{\odot}$ (see, e.g. \cite{Eckart-etal:2017:FOP:}). Thus, we get for electrons at the Galactic centre 
\beq \label{estimation-BBsgra}
\BB_{\rm SgrA^* (~e^-)} \approx \frac{|e| B G M}{2 m_e c^4} \approx 1.8 \times 10^{10}.
\eeq
For protons, the values of $\BB_{\rm SgrA^* (~p^+)} $ is of the order of $10^7$. In Table~\ref{Tab2} we show the values of $\BB$ for electron, proton, iron atom (without one electron), and a dust particle (without one electron) for various values of the MF strength and the three BH types (stellar mass, supermassive, and Galactic centre BH). Another robust way of estimating the MF in nearby the horizon region could be from observed relativistic jet, which is believed to be launched by the Blandford-Znajek mechanism \cite{Ripperda-etal:2022:ApJ:}.




\begin{table}[]
\begin{center}
\begin{tabular}{l  l  l  l  l}
\hline
 B [Gs]  &  \quad$\cb$ & \quad$k\cb$ & \quad$k\cb^2$ \\	
\hline
 $10^{15.8}$ & $\sim10^{19}$ & $\sim1$ & $\sim10^{19}$ \\ 
 $10^{12}$ & $\sim10^{15}$ & $\sim10^{-4}$ & $\sim10^{11}$\\  
 $10^8$ & $\sim10^{11}$ & $\sim10^{-8}$ & $\sim 10^{3}$ \\
$10^{6.4}$ & $\sim10^{9}$ & $\sim10^{-9}$ & $\sim1$ \\
 $10^4$ & $\sim10^{7}$ & $\sim 10^{-12}$ & $\sim10^{-5}$ \\ 
 $10^0$ & $\sim10^{3}$ & $\sim10^{-16}$ & $\sim 10^{-13}$ \\
 $10^{-2.9}$ & $\sim1$ & $\sim 10^{-19}$ & $\sim10^{-19}$ \\
 $10^{-5}$ & $\sim10^{-2}$ & $\sim 10^{-21}$ & $\sim10^{-23}$\\ 
\hline
\end{tabular}
\caption{Magnitudes of different forces acting on radiating charged particle in the dimensionless form of Eq. (\ref{curradforce}) for different values of magnetic field strengths, compared to the gravitational force of the order of $\sim1$. The estimates correspond to a relativistic electron in vicinity of a stellar mass BH with $M= 10~{M}_{\odot}$. 
\label{TabForce}}
\end{center}
\end{table}


\begin{table}[]
\begin{center}
\begin{tabular}{l  l  l  l  }
\hline
B [Gs]  &  $\tau_{\rm RR}^{e-}$ [s[  &  $\tau_{\rm RR}^{p+}$ [s] &  $\tau_{\rm RR}^{Fe+}$ [s] \\	
\hline
$10^{12}$ & $10^{-16}$ & $10^{-6}$ & $10^{-1}$ \\	
$10^{8}$ & $10^{-8}$ & $10^{2}$ & $10^{7}$ \\		
$10^{4}$ & $1$ & $10^{10}$ & $10^{15}$ \\	
$10^{0}$ & $10^{8}$ & $10^{18}$ & $10^{23}$ \\
$10^{-5}$ & $10^{18}$ & $10^{28}$ & $10^{33}$ \\		
\hline
\end{tabular}
\caption{
Typical cooling times of electron $\tau_e$, proton $\tau_p$, and ionized iron atom (without one electron) for different MF strengths. For charged dust particles, the synchrotron cooling is irrelevant. 
\label{Tab3}
}
\end{center}
\end{table}

RR force acting on a charged particle is represented by the dimensionless parameter $k$ which has the form
\beq \label{k-param}
k = \frac{2}{3} \frac{q^2}{m G M}.
\eeq
The value of parameter $k$ is much lower than that of $\BB$. For electron orbiting stellar mass and SMBH we have, respectively
\bea \label{estimation-k1}
k_{\rm BH} \sim 10^{-19} \quad {\rm for} \,\,\, M = 10 M_{\odot},  \\
k_{\rm SMBH } \sim 10^{-27} \quad {\rm for} \,\,\, M = 10^9 M_{\odot}.
\eea
For electron around Sgr A*, one gets $k_{\rm SgrA^*} \sim 10^{-25}$. For protons, the values of $k$ parameter is lower by the factor of $m_p/m_e \approx 1836$, as in the case with $\BB$. Despite the weakness of parameter $k$ as compared to $\BB$, it enters into the equations for ultrarelativistic particles as $k\BB^2$, which can make the effect of the RR force considerably large. These values for RR have been estimated for Sgr~A* and M87  in \cite{Ripperda-etal:2022:ApJ:} and in more details for M87 in \cite{Hakobyan-etal:2023:ApJ:}.

The motion of radiating charged particle is described by Eq. (\ref{curradforce}), containing several terms of different magnitudes. Let us estimate the contribution of each of the term in the equation separately and compare the individual terms with the  gravitational force. The left hand side of (\ref{curradforce})  is purely gravitational term given by the Christoffel symbols, which can be taken in the dimensionless units to be $\sim1$. On the right hand side of (\ref{curradforce}) one can distinguish the LF $\sim\cb$ and the contribution due to the RR consisting from the terms, which are proportional to $\sim k\cb$ and $\sim k^2\cb^2$. 
Representative values for all four terms forces are given in Table~\ref{TabForce}. Here we assume the particles to be relativistic $v\sim1$, which allows us to use the magnitude estimates for distinguishing different forces of particle dynamics, although the exact values of individual forces at given point may differ. 
Large values of $\BB$ in astrophysical settings suggest that the effects of EM interaction and RR can not be neglected since the LF and RR can be much stronger than gravity, but RR can overcome LF only for particles with large specific charge (electron) and for very strong MF of the order of $B>10^{15}$~Gs.

The RR timescale can be estimated from the following equation \cite{Tur-etal:2018:APJ:}
\beq 
\tau_{\rm RR} \approx \frac32 \frac{m^3 c^5}{f(r) q^4 B^2}, 
\eeq 
where $f(r)$ is a lapse function. Typical synchrotron RR timescales of particles in various MFs are given in Table~\ref{Tab3}. For the Galactic centre SMBH, the electron cooling timescale is of the order of $10^4$~s, while orbiting timescale at inISCO is  $\sim 10^3$~s. In general, for ions the cooling timescale is shorter than in case of electrons by the factor of $(m_p/m_e)^{3} \sim  10^{10}$. As one can see from the Table~\ref{Tab3}, the synchrotron radiation losses of ions and heavier particles are relevant in the presence of relatively strong MFs, i.e., much stronger than those corresponding to BHs, e.g., in the vicinity of neutron stars. 


\begin{table}[]
\begin{center}
\begin{tabular}{c  l  l }
\hline
   &  B [Gs] & $\omega_{\rm L}$ [Hz] \\	
\hline
inter stellar medium & $10^{-5}$ & $10^{2}$ \\  
Earth's MF & $10^{0}$ & $10^{6}$\\
Galactic center SMBH & $10^{1}$ & $10^{8}$\\ 
typical AGN & $10^{4}$ &  $10^{11}$\\
neutron star & $10^{12}$ & $10^{19}$\\
\hline
\end{tabular}
\caption{Magnitudes of some relevant astrophysical MFs and their related Larmor frequencies. \label{TabLarmor}
}
\end{center}
\end{table}

Fundamental frequencies for oscillating charged particle derived analytically in previous section can be relevant in astrophysical settings, e.g. in relation to the 
modulation of intensity flux (light curve) in quasi-periodic oscillations \cite{Kol-Stu-Tur:2015:CLAQG:, Kol-Tur-Stu:2017:EPJC:}, observed in many BH systems or in the model of the Galactic centre flares components  \cite{Tur-etal:2020:Flares:}. Fundamental frequencies can also be closely related to frequencies of radiated EM radiation \cite{Bre-etal:1973:PRD:}, hence they manifest themselves in produced EM spectrum and EM polarization. For strong MFs, the spectrum will be dominated by the Larmor frequency, estimated in Table~\ref{TabLarmor} for some typical astrophysical MFs. As one can see that for typical AGNs ($10^4$~Gs) the Larmor frequency is of the order of hundreds of GHz, similar to the observational frequency of the Event Horizon Telescope (230~GHz).


The synchrotron radiation losses of charged particles can be relevant in a plasma surrounding BH, if the cooling timescale is shorter or comparable with the mean free time between collisions of particles in plasma. The later can be estimated from the equation \cite{1981phki.book.....L}
\beq 
\tau_{ee} \approx \frac{m_e^2 (3 k_B T_e / m_e)^{3/2}}{\pi e^4 n_e 8 {\rm ln}\Lambda },
\eeq 
where ${\rm ln} \Lambda$ is a Coulomb logarithm, given by 
\beq 
{\rm ln} \Lambda  =  {\rm ln} \frac{3}{2e^3} \frac{(k_B T_e)^{3/2}}{\sqrt{\pi n_e}}.
\eeq 
Note that $\tau_{ee}$ represents the collisional timescale between electron-electron collisions in an electron-proton plasma. For proton-proton and electron-proton collisions at the same temperature, the timescales are typically longer. The representative estimates of collision timescales for given densities and temperatures of plasma are given in Table~\ref{Tab4}. 

It is interesting to estimate the electron collision timescales for the environment of the Galactic centre SMBH. A particle density in a plasma surrounding Sgr~A* can be obtained from the observed flares properties \cite{Tur-etal:2020:Flares:}, estimated to be of the order of $n \sim 10^{7\pm 1}$cm$^{-3}$. Modeling the emission of simultaneous NIR and X-ray flares from Sgr~A* the Lorentz factor of electrons are estimated to be of the order of $\gamma_e \sim 10^3$, which corresponds to the plasma with electron temperature $T_e \sim 10^9$K. Then, the electron-electron collision timescale is $\tau_{ee}$(Sgr~A*)$=10^{4\pm 1}$s. Note that this estimate is based on the flare observations, i.e., considered denser regions of the Galactic centre compared to its average surrounding. 
Such estimate depends on a number of assumptions - it may very well be that the density in a flaring region is lower, as long as the magnetization is high enough to power particles to Lorentz factors of $\sim 10^3$ \cite{Zhdankin-etal:2023:XXX}.
At larger spatial scales, averaging the plasma environment, the collision timescale is less by a few orders of magnitude \cite{2018MNRAS.480.4408Z}. Thus, the plasma surrounding Galactic centre can be considered as collisionless, so the considerations and estimates given in the current paper are fully applicable. In principle, there is more magnetic energy available in dilute (high sigma) regions, so flares could be powered from there. There are simulations showing that flares from Sgr~A* can be powered from magnetized regions, see \cite{Ripperda-etal:2022:ApJ:,Zhdankin-etal:2023:XXX}. This however does not change the main conclusion that the plasma is collisionless.

\begin{table}[]
\begin{center}
\begin{tabular}{l  l  l  l  }
\hline
$n$ [cm$^{-3}$]  &  $T$ [K]  &  $\tau_{ee}$ [s] &  $\tau_{ii}$ [s] \\	
\hline
$10^{14}$ & $10^{10}$ & $10^{-1}$ & $1$ \\	
$10^{14}$ & $10^{6}$ & $10^{-7}$ & $10^{-5}$ \\			
$10^{10}$ & $10^{10}$ & $10^{3}$ & $10^{4}$ \\	
$10^{10}$ & $10^{6}$ & $10^{-3}$ & $10^{-1}$ \\
$10^{6}$ & $10^{10}$ & $10^{7}$ & $10^{8}$ \\
$10^{6}$ & $10^{6}$ & $10$ & $10^{3}$ \\
\hline
\end{tabular}
\caption{
Collision timescale of electron-electron $\tau_{ee}$ and ion-ion $\tau_{ee}$ collisions in a plasma of given particle density and temperature. 
The plasma is considered quasi-neutral with equal temperatures of the components. The values are rounded to the orders of magnitude. 
\label{Tab4}
}
\end{center}
\end{table}



\section{Conclusions} \label{kecy}

The effects of MFs around BHs cannot be neglected in many astrophysical settings. If the specific charge of a test particle $q/m$ is large enough, as in the case of elementary particles, even a weak MF can significantly influence the particle dynamics, the position of the inner edge of an ionized Keplerian accretion disk, or the epicyclic frequencies for motion around BH. In this article we have tested a simple but realistic model for BH magnetosphere given by the parabolic MF and explored influence of RR damping force on chaotic charge particle dynamics. We also briefly discussed the astrophysical relevance of the presented theoretical model. 

We studied the parabolic MF configuration around BH in general form for different values of the inclination parameter $w$, limiting to the magnetic split-monopole solution for $w=0$ as a special case. The parabolic configuration of MF is supported by observations and GRMHD numerical experiments \cite{Nak-etal:2018:APJ:}. Generation of such MF geometry would require the presence of electric current sheets at the equatorial plane provided by ongoing accretion processes. 
The accretion disk is located in equatorial plane and outside of the equatorial plane, the matter density is expected to be much lower. Therefore, in discrete particle approach used in the present paper, the existence of charged particle off-equatorial orbits become especially important. 

We have analysed stability of circular orbits of charged particles and found that the off-equatorial stable circular orbits may exist only in the case of attractive LF ($\cb<0$), while the stable circular orbits at the equatorial plane can be observed only in the case of repulsive LF ($\cb>0$). The effective potential minima situated at the off-equatorial plane get closer to the $z$-axis when the MF becomes stronger. In the case of the in-equatorial plane, the radius of inISCO of the charged particle decreases as the MF strength increases, approaching the BH horizon $r_{\rm inISCO} = 2$ at large values of the magnetic parameter $\cb$. In the case of the split magnetic monopole, the inISCO (or offISCO) remains at a constant radius $r_{\rm inISCO}=6$. 

We examined the regular and chaotic behaviour of the motion of charged particles with the help of PSs, which is a good tool illustrating the difference between regular and chaotic dynamics. We have found that charged particle motion around BH in parabolic MF can be chaotic under certain initial conditions. This result is in contrast with the neutral particle case or the case with the monopole MF configuration, where due to the symmetry of the EM and gravitational fields, chaos does not occur and the motion is always regular \cite{Kol-Bar-Jur:2019:RAGtime:}. We have also compared the change of the chaoticity of trajectories when RR force in included in the equations of motion and found that the RR acts as a damping force, creating new attractors in phase space, which are situated on circular orbit at the equatorial plane of the BH. We have found that the off-equatorial plane orbits are always unstable when the radiation is taken into account, since the RR force tends to shift the charged particle orbits to their guiding centers, which coincides with the position of the BH. 
The RR force has damping character of the influence on the charged particle dynamics and the final stage of the radiating particle motion is eventually a collapse into the BH if the integration time is long enough. The particle loses its energy and angular momentum due to RR cooling, which leads to spiraling down to the BH. 
For positive value of magnetic parameter $\cb>0$, an oscillating charged particle experiencing the RR force radiates its oscillations away and settles down to the quasi-circular orbits, before eventually falling down into the BH.

We have also studied the influence of the RR force on the structure of a thin Keplerian accretion disk orbiting  Schwarzschild BH in external parabolic MF. The effect of RR can be observed even for a relatively small values of $\cb$. For $|\cb|\gg1$, one can observe oscillations in the disk dynamics, for $|\cb|\sim1$, the disk trajectories enter into the turbulent chaotic regime and for $|\cb|\ll1$, the charged particle trajectories  stick to MF lines. Due to RR force, disk loses energy and angular momentum and particle will end up into BH. The time scale of the destruction of the Keplerian disk by radiation strongly depends on the strength of MF and radiation parameter $k\cb$.

We have calculated the fundamental frequencies of a  charged particle and produced EM spectrum, which can be useful in determining time scales of various astrophysical processes around BH and fitting observed data. Test particle frequencies are fundamental in the sense that they will appear in any more advanced and complicated model for BH neighborhood description, like GRMHD simulation as a zero approximation \cite{Mis-Klu-Fra:2019:MNRAS:}. It is worthwhile mentioning that our analytical solutions are very important to test the codes that evolve particles (e.g. GRPIC), for example in the codes of \cite{Parfrey-etal:2019:PhRvL:,Bacc-etal:2019:ApJS:,Bacchini-etal:2020:ApJS:}.
One example where particle frequencies could be crucial is the quasi-periodic oscillations of the X-ray power density observed in quasars and microquasars. Fitting these quasi-periodic oscillation frequencies in timing spectra can help to verify BH magnetosphere model \cite{Kol-Tur-Stu:2017:EPJC:,Sha-etal:2021:EPJC:}. Another observational window for BH accretion processes could be non-thermal synchrotron emission produced by charged particles moving in the BH magnetosphere \cite{Fro-etal:2021:ARXIV:}. One charged particle dynamics and its oscillations around effective potential minima could be used to construct synthetic charged particle synchrotron spectra \cite{Sho:2015:PRD:,Crinquand-etal:2022:PhRvL:,Hakobyan-etal:2023:ApJ:}

We have also estimated the most relevant parameters of the discussed model in astrophysical situations and found the limits on the relevance of the effects of the LF and RR for different particle types and BH systems. For that, we gave the estimates of orbital, radiation, and collisional timescales of particles. We have shown that applied to the Galactic centre BH the plasma environment can be considered as collisionless and test particle approach can be used there. We also discussed the role of MF in fundamental frequencies, observable in relation to the so-called quasi-periodic oscillations. 

Among the possible extensions of the presented parabolic magnetosphere model, which could be interesting to explore in future are the calculation of the radiated EM spectra and polarization, and inclusion of the BH rotation into the model. In relation to the existence of the off-equatorial orbits found in this paper, it would also be interesting to explore, whether the stable off-equatorial structures, existing in discrete particle approach can be found in numerical GRMHD or PIC simulations. This is a prediction that needs to be tested.

\section*{Acknowledgments}


The work is supported by the Research Centre for Theoretical Physics and Astrophysics, Institute of Physics, Silesian University in Opava and the GA{\v C}R \mbox{23-07043S} project.



\def\prc{Phys. Rev. C}
\def\pre{Phys. Rev. E}
\def\prd{Phys. Rev. D}
\def\jcap{Journal of Cosmology and Astroparticle Physics}
\def\apss{Astrophysics and Space Science}
\def\mnras{Monthly Notices of the Royal Astronomical Society}
\def\apj{The Astrophysical Journal}
\def\aap{Astronomy and Astrophysics}
\def\actaa{Acta Astronomica}
\def\pasj{Publications of the Astronomical Society of Japan}
\def\apjl{Astrophysical Journal Letters}
\def\pasa{Publications Astronomical Society of Australia}
\def\nat{Nature}
\def\physrep{Physics Reports}
\def\araa{Annual Review of Astronomy and Astrophysics}
\def\apjs{The Astrophysical Journal Supplement}
\def\prl{Physical Review Letters
}
\def\aapr{The Astronomy and Astrophysics Review}
\def\procspie{Proceedings of the SPIE}

\bibliographystyle{spphys} 

\bibliography{reference}

\end{document}